\documentclass[preprint,aps,tightenlines,showkeys,nofootinbib,superscriptaddress]{revtex4}
\usepackage{latexsym,amssymb,amstext,amsmath}
\usepackage[dvips]{graphicx}
\usepackage[dvips]{color}
\newcommand{\beq}{\begin{eqnarray}}
\newcommand{\eeq}{\end{eqnarray}}
\newcommand{\bea}{\begin{eqnarray*}}
\newcommand{\eea}{\end{eqnarray*}}
\newcommand{\eq}{eqnarray}

\newcommand{\al}{{\alpha}}
\newcommand{\be}{{\beta}}
\newcommand{\ci}{\cite}
\newcommand{\ga}{{\gamma}}

\newcommand{\De}{\Delta}

\newcommand{\la}{{\lambda}}
\newcommand{\La}{{\Lambda}}

\newcommand{\si}{{\sigma}}

\newcommand{\ka}{{\kappa}}
\newcommand{\om}{{\omega}}
\newcommand{\Om}{{\Omega}}

\newcommand{\no}{{\nonumber}}
\newcommand{\f}{\frac}

\newcommand{\Ho}{Ho\v{r}ava}

\begin{document}

\preprint{arXiv:2108.07986v3b [hep-th]}
\title{Tests of Standard Cosmology in Ho\v{r}ava Gravity{,} Bayesian Evidence for a Closed Universe{, and} the Hubble Tension
}
\author{Nils A. Nilsson\footnote{E-mail address: albin.nilsson@ncbj.gov.pl, On leave from
\it{Rutherford Appleton Laboratory (RAL), Harwell Campus, Didcot,
OX11 0QX, United Kingdom;
National Centre for Nuclear Research, Pasteura 7,
00-293, Warsaw, Poland}}
and Mu-In Park\footnote{E-mail address: muinpark@gmail.com, Corresponding author}}
\affiliation{ Center for Quantum Spacetime, Sogang University,
Seoul, 121-742, Korea }
\date{\today}

\begin{abstract}
{We consider some background tests of standard cosmology in the context of Ho\v{r}ava gravity with different scaling dimensions for space and time, which has been proposed as a renormalizable, higher-derivative, Lorentz-violating quantum gravity model without ghost problems. We obtain the ``very strong"
and ``strong" Bayesian evidences
for our two cosmology models {\bf A} and {\bf B}, respectively, depending on the
choice of parametrization based on \Ho~gravity, against the standard,
spatially-flat, LCDM cosmology model based on general relativity. An MCMC analysis
with the observational data, including BAO, shows (a) preference of a
{\it closed}
universe with the curvature density parameter
$\Omega_k=-0.005\pm 0.001$, $-0.004^{+0.003}_{-0.001}$, (b) reduction of
the Hubble tension with the Hubble constant
$H_0=71.4^{+1.2}_{-0.9}$, $69.5^{+1.6}_{-0.9}~ {\rm km s^{-1} Mpc^{-1}}$ for
the models {\bf A}, {\bf B},
{respectively, and also (c) a positive result on the discordance problem}.
We comment on some possible further improvements for the ``cosmic-tension problem" by
considering the more complete early-universe physics,
based on the Lorentz-violating standard model with anisotropic space-time scaling, consistently with Ho\v{r}ava gravity, as well as the observational data which are properly adopted for the closed universe.}
\end{abstract}

\keywords{Standard Cosmology Tests, Ho\v{r}ava Gravity Cosmology, Hubble Tension, Non-flat Universe}

\maketitle

\newpage
Standard cosmology, which is usually formulated as the Lambda Cold Dark Matter (LCDM) model, is based on general relativity (GR) with a positive cosmological constant \cite{Frie:1922,Lema:1927,Ries:1998} and has been quite successful in describing the observational data \cite{WMAP:2010}. However, with the increased accuracy of data, the significant deviations from LCDM are becoming clearer \cite{Ries:2018,Ries:2019,Asga:2019}. In particular, regarding the recent discrepancies of the Hubble constant from the Cosmic Microwave Background (CMB) data and the direct (local) measurements at the lower redshift, which corresponds to the mismatches between the early and late universes, there have been various proposals (for recent reviews, see \cite{DiVa:2021,Peri:2021,Scho:2021}) but it is still a challenging problem to
find a resolution at the fundamental level.

{On the other hand, if our universe was created from the
Big Bang, we need a {\it quantum gravity} to describe the
early universe or later
space-time. But, it has been well known that a renormalizable
quantum gravity can not be realized in GR or its (relativistic)
higher-derivative generalizations, due to the ghost problem \ci{Stelle:1976}.
Recently, Ho\v{r}ava has proposed a renormalizable,
higher-derivative, {\it Lorentz-violating} quantum gravity model without the ghost problem, due to the high-energy (UV) Lorentz-symmetry violations
from the different scaling dimensions for space and time
{\it \`{a} la} Lifshitz, DeWitt, and Ho\v{r}ava
\cite{Lifs:1941,DeWi:1967,Hora:2009}. In the last 12 years, there have
been many works on its various aspects (see \ci{Wang:2017} for a review and
extensive literatures). Theoretically, there are still several
fundamental issues, like the full/complete symmetry structure,
full dynamical degrees of freedom,
renormalizability, and the very meaning of black holes and Hawking
radiation, etc. (see \ci{Deve:2021} for 
related discussions and
their current status). However, phenomenologically, Ho\v{r}ava gravity
is one of the (viable) modified gravity theories and it can be tested from
astrophysical or cosmological observations \ci{Olmo:2019}. In particular,
from the recent detections of gravitational waves from black
holes/neutron stars, the importance of quantum gravity and its
{\it non-GR} behaviors is increasing
 \ci{LIGO:2021}.
 There are some interesting 
 results on {\it testable}
 Ho\v{r}ava gravity effects, such as the increased  maximum mass of neutron stars
 \ci{Kim:2018},
 reduced light deflection \ci{Liu:2010} and black hole
 shadow \ci{Li:2021}, etc. There are also some constraints on its
 low-energy limit or Einstein-Aether theory from
 astrophysical data \ci{Emir:2017,Gong:2018,Khod:2020,Gupt:2021} or
 cosmological data with the ``assumed" spatially-{\it flat} universe in standard cosmology \ci{Frus:2015,Frus:2020}. But still, for a renormalizable Ho\v{r}ava gravity with the desired higher-derivative terms, there are no {\it systematic} and {\it significant} constraints from the observational data.
}

In this paper, we test the spatially {\it non-flat} universe
in standard cosmology in the context of {Ho\v{r}ava gravity}.
A peculiar property of the cosmology based on Ho\v{r}ava gravity is that the spatially-curved universe may be more
``natural" due to contributions from higher spatial derivatives.
For the spatially-flat universe, on the other hand, the usual FLRW {\it background} cosmology
in Ho\v{r}ava gravity \cite{Frie:1922,Lema:1927} is the same as in GR and hence
there are no observable differences in the data analysis which means
the same (background) tensions in LCDM model. In other words, Ho\v{r}ava gravity
is a ``natural laboratory" for the tests of the spatially non-flat universe in
standard cosmology.

Recently, it has been found \cite{DiVa:2019,Hand:2019,DiVa:2020,eBOS:2020}
that the tensions get worse in LCDM as $-\Omega_k$ increases, {\it i.e},
a rather lower value of Hubble constant for a closed universe ($\Omega_k<0$),
which is preferred in the recent Planck CMB data \cite{Planck:2018}, {without being combined with lensing and Baryon Acoustic Oscillations (BAO)}. In this
paper, we show that the situation in Ho\v{r}ava gravity is the opposite,
{\it i.e.}, tensions get better with an increasing $-\Omega_k$, due
to some non-linear corrections from (spatial) higher-derivative terms.
From an MCMC analysis with the observational data, including BAO, we
obtain (a) preference of a {\it closed} universe
and (b) reduction of the Hubble tension
for our two \Ho-gravity based cosmological models {\bf A} and {\bf B}, depending on the choice of parametrization, with ``very strong" and ``strong" Bayesian evidences,
respectively, against the standard, spatially-flat, LCDM cosmological model.

To this ends, we consider the ADM {(Arnowitt-Deser-Misner \cite{Arno:1962})} metric
\begin{\eq}
ds^2=-N^2 c^2 d t^2
+g_{ij}\left(dx^i+N^i dt\right)\left(dx^j+N^j dt\right)\,
\end{\eq}
and the Ho\v{r}ava gravity action with $z=3$, {\it \`{a} la} Lifshitz, DeWitt, and Ho\v{r}ava \cite{Lifs:1941,DeWi:1967,Hora:2009},
given by (up to boundary terms)
\begin{\eq}
\label{HL action}
S_\mathrm{g} &=& \int d t d^3 x \sqrt{g} N \left[ \frac{2}{\kappa^2}
\left(K_{ij}K^{ij} - \lambda K^2 \right) - {\cal V} \right], \\
-{\cal V}&=& \sigma+ \xi R + \alpha_1 R^2+ \alpha_2 R_{ij}R^{ij}
+\alpha_3 \frac{\epsilon^{ijk}}{\sqrt{g}}R_{il}\nabla_jR^l{}_k \no \\
 &+& \alpha_4 \nabla_{i}R_{jk} \nabla^{i}{R}^{jk}
+\al_5 \nabla_{i}R_{jk}\nabla^{j} {R}^{ik}
+\al_6 \nabla_{i}R\nabla^{i}R , 
\end{\eq}
which is viable \footnote{At the perturbation level, even for the
spatially-flat background, \Ho~gravity produces
notable differences from GR.
In order to obtain a (nearly) scale-invariant CMB power spectrum for the
spatially-flat universe in Ho\v{r}ava gravity, where the ``inflation without inflation era" is possible \cite{Muko:2009,Gesh:2011},
we need a {\it proper}
form of the six-derivative (UV) terms which break the detailed balance condition
in UV as in the action (\ref{HL action}) \cite{Shin:2017}. It would be interesting to see
the curvature-induced effect on the scale-invariant power spectrum.
 However, in this paper, we consider the action (\ref{HL action}) without any UV conditions for the generality of our approach.} \cite{Shin:2017} and power-counting renormalizable \cite{Viss:2009},
with the extrinsic curvature
\begin{\eq}
K_{ij}=\frac{1}{2N}\left(\dot{g_{ij}}
-\nabla_i N_j-\nabla_jN_i\right)\
\end{\eq}
(the dot $(\dot{~})$ denotes a time derivative),
the  Ricci tensor $R_{ij}$ of the  (Euclidean) three-geometry, their corresponding traces $K=g_{ij} K^{ij}, R=g_{ij} R^{ij}$,
coupling constants \footnote{One might consider
extension terms which depend on $a_i=\partial_i N/N$ and $\nabla_j a_i$ generally \cite{Blas:2009}. However, since these terms {considerably} affect the IR physics
{compared to those in}
 the standard action (\ref{HL action}) {\cite{Deve:2018,ONea:2020}}, we do not consider those terms in this paper by assuming that the IR physics is well described by GR. This implies that the gravity probe $E_G$ approaches the GR prediction in the current epoch, {\it i.e.,} low $z$, though currently it would not be tested due to large statistical errors \cite{Zhan:2020}.} $\kappa,\lambda,\xi,\alpha_i$, {and a cosmological constant parameter $\si$}.

In order to study standard cosmology for the \Ho~gravity action
(\ref{HL action}), we consider the homogeneous and isotropic Friedmann-Lemaitre-Robertson-Walker (FLRW) metric ansatz
\begin{\eq}
ds^2=-c^2dt^2+a^2(t)\left[\frac{dr^2}{1-kr^2/R_0^2}+r^2\left(d\theta^2+\sin^2\theta
d\phi^2\right)\right]
\end{\eq}
with the (spatial) curvature parameter $k=+1,0,-1$ for a closed, flat,
open universe, respectively, and the curvature radius $R_0$ in the current epoch $a(t_0) \equiv 1$. Assuming the perfect fluid form of matter contributions with energy density $\rho$ and
pressure $p$, we obtain the Friedmann equations as
\begin{\eq}
H^2&=&
\frac{\kappa^2}{6(3\lambda-1)}
\left[\rho + \frac{3\kappa^2\mu^2}{8(3\lambda-1)} \left( \f{
k^2}{R_0^4 a^4}+ \f{2 k (\om-\La_W )}{R_0^2 a^2}+ \La_W^2 \right) \right] ,  \label{F1}\\
\dot{H}+H^2&=&\f{\ddot{a}}{a}=\frac{-\kappa^2}{6(3\lambda-1)} \left[\f{1}{2}
(\rho+3 p) + \frac{3 \kappa^2\mu^2}{8(3\lambda-1)} \left( \f{
k^2}{R_0^4 a^4}- \La_W^2 \right) \right], \label{F2}
\end{\eq}
where we have used the conventional parametrization of the coupling constants $\sigma, \xi, \al_1, \al_2$ for the lower-derivative
terms \cite{Hora:2009,Park:2009}
\begin{\eq}
\label{parameters}
&&\sigma=\f{3 \kappa^2 \mu^2 \La_W^2}{8 (3 \la-1)},~
\xi=\f{\ka^2 \mu^2 (\om-\La_W)}{8 (3 \la-1)},~
\al_1=\f{\ka^2 \mu^2 (4 \la-1)}{32 (3 \la-1)},~
\al_2=-\f{\ka^2 \mu^2 }{8}
\end{\eq}
with {an IR-modification parameter $\om$} \cite{Park:2009}, $\mu^2>0 ~(<0)$ for a positive (negative) cosmological constant $(\sim \La_W)$, and the Hubble parameter $H(t) \equiv \dot{a}/a$. However, we take the coupling constants $\al_3,\cdots, \al_6$ for higher-derivative (UV) terms to be arbitrary so that a (nearly) scale-invariant CMB spectrum with respect to the background universe, as well as power-counting renormalizability, can be obtained \cite{Shin:2017} \footnote{For the cosmological perturbation around the spatially-flat FLRW background, the scale-invariant spectrum for the tensor modes depend only on the coupling $\al_4$, whereas the scalar mode depends on the combination $\widetilde{\al}_4 \equiv \al_4 +2 \al_5/3+8 \al_6/3$.}.
However, it is important to note that there are no contributions in the above Friedmann equations from the fifth and sixth-derivative UV terms in the \Ho~gravity action (\ref{HL action}) due to $R_{ij}={2 k}g_{ij}/{R_0^2 a^2(t)} ,~ K_{ij}=H(t) g_{ij}$, but only from the fourth-derivative terms, which leads to $k^2/a^4$ terms  \footnote{If we include {\it non-derivative} higher-curvature terms, like $R^3, R^2 R_{ij} R^{ij}$, etc., we have $a^{-6}$ terms as well, which correspond to
{\it stiff} matter \cite{Dutt:2009,Son:2010}. However, in this paper we do not consider those terms for simplicity.} in the Friedmann equations (\ref{F1}) and (\ref{F2}).
On the other hand, for the spatially-flat universe with $k=0$, all the
contributions from the higher-derivative terms disappear and we recover the same background cosmology as in GR, which means a return to the LCDM model with the Hubble constant tension. In this sense, Ho\v{r}ava gravity is a ``natural laboratory" for the tests of the spatially non-flat universe in standard cosmology.

Introducing
dust matter (non-relativistic baryonic matter and (non-baryonic) cold dark matter with $p_m=0$) and radiation
(ultra-relativistic matter with $p_r=\rho_r/3$), which satisfy the continuity equations
$\dot{\rho}_i+3 H (\dot{\rho}_i+p_i)=0 ~(i=m
, r
)$, we conveniently define the canonical density parameters at the current epoch $a_0=1$ as \footnote{We adopt the common convention $\Om_{i}$ for the current values and $\Om_{i} (a)$ for the fully time-dependent values.}
\begin{\eq}
\Omega_m \equiv \be \f{ \rho^0_m}{3H_0^2},~ \Omega_r\equiv \be \f{\rho^0_r}{3H_0^2},~ \Omega_k\equiv-\ga \f{k}{H_0^2R_0^2}, ~ \Omega_\Lambda\equiv\ga \f{\Lambda_W}{2H_0^2},~ \Omega_\omega\equiv\ga \f{\omega}{2H_0^2},
\end{\eq}
\\
where $\be \equiv  \kappa^2/2(3\lambda-1)$, $\ga \equiv \kappa^4 \mu^2 \La_W/8(3\lambda-1)^2$ are
positive constant parameters. Then, we can write the (first) Friedmann equation (\ref{F1}) as
\begin{equation}\label{H_eq}
    \left(\frac{H}{H_0}\right)^2 = \Omega_r a^{-4}+\Omega_m a^{-3}  + \Omega_k a^{-2} + \Omega_{\rm DE}(a),
\end{equation}
where we have introduced the (dynamical) dark-energy (DE) component as
\begin{equation}
    \Omega_{\rm DE}(a) \equiv \left(\frac{\Omega_k^2}{4\Omega_\Lambda}\right) a^{-4}
    -\left(\frac{\Omega_k \Omega_\omega }{\Omega_\Lambda}\right)a^{-2} + \Omega_\Lambda,
    \label{DE}
\end{equation}
which is defined as all the extra contributions to the first-three GR terms in (\ref{H_eq})
\cite{Park:2009}. Here, we note that this dynamical dark-energy component includes the
{\it dark radiation} (DR) and {\it dark curvature} (DC) components as
\begin{\eq}\label{DR}
    \Omega_{\rm DR}(a)\equiv \left(\frac{\Omega_k^2}{4\Omega_\Lambda}\right) a^{-4},~
   \Omega_{\rm DC}(a)\equiv
   -\left(\frac{\Omega_k \Omega_\omega }{\Omega_\Lambda}\right)a^{-2},
\end{\eq}
{which play the roles of the extra radiation and curvature terms,} respectively, as well as the usual cosmological constant component $\Omega_\Lambda$, so that
$\Omega_{\rm DE}(a)=\Omega_{\rm DR}(a)+\Omega_{\rm DC}(a)+\Omega_\Lambda$.

So far we have not assumed any specific information about the early universe. The whole
physics of early universe in the context of Ho\v{r}ava gravity would be quite different
from our known physics and needs to be revisited for a complete analysis. However, as in the standard cosmology,
we may introduce the phenomenological parametrization so that
all
the unknown (early-universe) physics can be taken into account.
For example, regarding 
the Big Bang Nucleosynthesis (BBN) at the decoupling
epoch $a_{\rm dec}=(1+z_{\rm dec})^{-1}=(1091)^{-1}$,
the {\it early dark
radiation} 
can be expressed as the contribution from the hypothetical excess $\Delta N_{\rm eff}$ in
the {standard model prediction of the} effective number of
{relativistic} species $N_{\rm eff} = 3.046$ as \cite{Stei:2012}
\begin{\eq}
\Om_{\rm DR}(a) &\equiv& \f{7}{8}\left( \f{4}{11} \right)^{4/3} \Delta N_{\rm eff} ~\Om_{\ga} a^{-4} \no \\
&=&0.13424~ \Delta N_{\rm eff}~ \Om_{r} a^{-4}, \label{BBN_DR_relation}
\end{\eq}
where we have used the present radiation density parameter for the standard model particles with
negligible masses (photon and three species of neutrinos),
$\Om_r=[1+ \f{7}{8}( \f{4}{11})^{4/3} N_{\rm eff}] ~\Om_{\ga}=\f{7}{8}( \f{4}{11})^{4/3} \cdot (0.13424)^{-1} \Om_{\ga}$
with the photon density $\Om_{\ga}=2.4730 \times 10^{-5} h^{-2}$ for
the present CMB temperature $T_0=2.7255 K$ \cite{Planck:2018} and $h \equiv H_0/100 {\rm~ km s^{-1} Mpc ^{-1}}$. If we use  the relation (\ref{BBN_DR_relation}) for the dark radiation formula (\ref{DR}) in \Ho~gravity, we can express the cosmological constant component $\Om_{\La}$ as
\begin{\eq}
\Om_\La=\f{\Om_k^2}{4 \cdot 0.13424~ \Delta N_{\rm eff}~ \Om_r} \label{CC_relation}
\end{\eq}
so that the Friedmann equation (\ref{H_eq}) can be written by $\Delta N_{\rm eff}$, instead of $\Om_{\La}$ \cite{Dutt:2009,Nils:2018}. Here, it is important to note that $\Delta N_{\rm eff}$ needs not to be an integer but can be an arbitrary and positive (negative) real number for a positive (negative) $\Om_{\La}$, {\it i.e.,} asymptotically {\it de Sitter} ({\it Anti de Sitter}) universe with a cosmological constant $\sim \La_W$. Moreover, we note that the relation (\ref{CC_relation}) implies an intriguing correlation between $\Om_k, \Om_r, \Delta N_{\rm eff}$, and $\Om_{\La}$, {which are otherwise unrelated.}

In this paper, we consider two models,
{\bf A} and
{\bf B}, depending on whether we implement (\ref{CC_relation}) or not, to see the effectiveness of the BBN-like parametrization in terms of $\Delta N_{\rm eff}$. Then, for model {\bf A}, the Friedmann equation (\ref{H_eq}) reduces to
\begin{\eq}
\label{H_eq:Model A}
    \left(\frac{H}{H_0}\right)^2 &=& (1+0.13424~ \Delta N_{\rm eff}) \Omega_r a^{-4}+\Omega_m a^{-3} \no \\ &+& \left[ \Omega_k - 4 \cdot 0.13424~ \Delta N_{\rm eff} \left(\f{\Om_{\om} \Om_r}{\Om_k}\right)\right] a^{-2}
    + \f{\Om_k^2}{4 \cdot 0.13424~ \Delta N_{\rm eff}~ \Om_r},
\end{\eq}
which is one of our main equations for comparing with cosmological data, with the assumption of non-vanishing $\Om_k$ and $\Delta N_{\rm eff}$ for the well-defined equation (\ref{H_eq:Model A}). Here we note that, considering $\Om_r$ is a function of $H_0$ as given above,
this model is described by five non-linearly coupled parameters
$H_0, \Om_m, \Om_k, \Om_\om, \Delta N_{\rm eff}$, in contrast to the three linear, decoupled parameters
$H_0, \Om_m, \Om_\La$ for the standard, spatially-flat (background) cosmology, LCDM \footnote{From the current-epoch constraint, {\it i.e.,} $H(t_0)\equiv H_0$ in (\ref{H_eq}),
the number of independent parameters reduces to
4 and 2 for our two models based Ho\v{r}ava gravity and the standard models
based on GR, respectively. In this paper, we conveniently take
$H_0, \Om_k, \Om_\om, \Delta N_{\rm eff}$(or $\Om_\La$) and
$H_0, \Om_k, \Om_\La$ for the models {\bf A}, {\bf B}, and (k)LCDM,
respectively, by considering $\Om_m$, as well as $\Om_r$, as the derived
(dependent) parameter. However, we need to introduce the baryonic parameter
$\Om_b$ as an independent parameter when we analyze CMB and BAO data later
so that only the (non-baryonic) dark matter sector in $\Om_m$ is the derived
parameter.}.

On the other hand, for model {\bf B}, the Friedmann equation (\ref{H_eq}) is simply given by
\begin{equation}\label{Model_B}
    \left(\frac{H}{H_0}\right)^2 = \left( \Omega_r + \frac{\Omega_k^2}{4\Omega_\Lambda} \right) a^{-4}+\Omega_m a^{-3}  + \left(1-\frac{\Omega_\omega }{\Omega_\Lambda}\right) \Omega_k a^{-2} + \Omega_{\La}
\end{equation}
without using the relation (\ref{CC_relation}) from the BBN-inspired formula
(\ref{BBN_DR_relation}) for the dark radiation in Ho\v{r}ava gravity. This model
has two (non-linearly) coupled
parameters $\Om_k$ and $\Om_\om$,
in addition to those of standard flat cosmology, $H_0, \Om_\La$, with assuming
a non-vanishing $\Om_\La$ for the well-defined equation (\ref{Model_B}). We also
consider the spatially flat $(k=0)$ LCDM as well as non-flat $(k \neq 0)$ kLCDM
for
comparison.

We probe our universe by using a Markov-Chain Monte Carlo (MCMC) method for
the four models,
{\bf A}, {\bf B}, LCDM, kLCDM and determine the independent parameters with the statistical inferences for the models. We use the Metropolis-Hastings algorithm \footnote{We use the Metropolis-Hastings algorithm
at the background level
since there is no known analysis
of
the
cosmological perturbations
for a non-flat universe, contrary to the flat universe \cite{Shin:2017}, due to
computational complications in Ho\v{r}ava gravity.
 However, the algorithm has the limitation in our case, though not a general feature, that does not show the
 proper $\chi^2$ values for the {\it separate} data sets, while it shows
 the convergent results for the {\it whole} data sets.
 } \cite{Robe:2015} for the posterior parameter distributions, the statistical method in \cite{Dunk:2004} for convergence tests of the MCMC chains, and the public code GetDist \cite{Lewi:2019} for the visualization of the results. The cosmological data sets we consider (for the details, see Appendix {\bf A}) are CMB (Planck 2018 \cite{Zhai:2018}), BAO (SDSS-BOSS \cite{BOSS:2016}, SDSS-IV \cite{Ata:2017}, Lyman-$\alpha$ forest \cite{deSa:2019},
and WiggleZ
\cite{Blak:2012}),
SNe Ia (Pantheon
\cite{Scol:2017}), GRBs (Mayflower
\cite{Liu:2014}), Lensed Quasars (H0liCOW \cite{Wong:2019}), and Cosmic
Chronometers (CC)
\cite{More:2015}.
{Here, it is important that we need to include CMB lensing data in combination with BAO or SNe Ia in order to test the curvature of the universe \cite{Efst:2020,Gonz:2021}.}

Our main results are shown in Table 1 and their essentials are plotted in Fig. 1 (for the full plots, see Appendix {\bf B}). Some noticeable results are as follows:
\\
\renewcommand{\arraystretch}{1.0}
\begin{table}[h!]
\begin{tabular}{|l|c|c|c|c|}
\hline
Model Parameters & Model {\bf A} & Model {\bf B} & kLCDM & LCDM\\
\hline
$\Omega_b h^2$ & $0.0227 \pm 0.0001$ & $0.0227 \pm 0.0001$ & $0.0226 \pm 0.0001$ & $0.0225 \pm 0.0001$ \\
$\Omega_m$ & $\mathbf{0.307\pm 0.004}$ & $\mathbf{0.306^{+0.005}_{-0.006}}$ & $\mathbf{0.302 \pm 0.005}$ & $\mathbf{0.305 \pm 0.004}$ \\
$\Omega_r \cdot 10^{5}$ & $\mathbf{8.20^{+0.22}_{-0.27}}$ & $\mathbf{8.64^{+0.23}_{-0.38}}$ & $\mathbf{8.93^{+0.14}_{-0.13}}$ & $\mathbf{8.94^{+0.08}_{-0.08}}$ \\
$\Omega_k$ & $-0.005\pm {0.001}$ & $-0.004^{+0.003}_{-0.001}$ & $-0.001 \pm 0.002$ & - \\
$\Omega_\Lambda$ & $\mathbf{0.70\pm 0.01}$ & $0.695\pm 0.005$ & $0.699 \pm 0.004$ & $0.695 \pm 0.004$ \\
$H_0$ [km s$^{-1}$ Mpc$^{-1}$] & $71.38^{+1.19}_{-0.93}$ & $69.53^{+1.57}_{-0.91}$ & $68.41^{+0.52}_{-0.51}$ & $68.36^{+0.32}_{-0.30}$ \\
$\Omega_\omega$ & $-0.75^{+0.46}_{-0.24}$ & $0.34^{+1.15}_{-0.31}$ & - & - \\
$\Delta N_{\rm eff}$ & $0.87^{+0.28}_{-0.26}$ & - & - & - \\
\hline
$\chi^2_{\rm min}$ & $1143.6$ & $1150.1$ & $1155.0$ & $1157.9$\\
$\Delta \chi^2_{\rm min}$ & $-14.3$ & $-7.8$ & $-2.9$ & $0$\\
$ \ln{E}$ & $-573.8
$ & $-576.9
$ & $-578.9
$ & $-580.0
$\\
$\ln{B_{ij}}$ & $+6.2
$ & $+3.1
$ & $+1.1
$ & $0$ \\
\hline
\end{tabular}
\caption{Constraints at $1 \sigma~ (68 \%)$ CL errors on the cosmological parameters
for our two cosmology models
{\bf A}, {\bf B}, based on Ho\v{r}ava gravity,
and the standard cosmology models, spatially-flat $(k=0)$ LCDM and spatially-non-flat ($k \neq 0$) kLCDM, based on GR. In the bottom lines, we show $\chi^2_{\rm min}$ for the best-fit values of parameters, the (logarithmic) Bayesian evidence $\ln{E}$, and the Bayes factor $\ln{B}_{ij}$ with respects to LCDM. $\Delta \chi^2_{\rm min}$ represents the difference of $\chi^2_{\rm min}$ from those of LCDM. The {\bf bold-faced} quantities represent the derived quantities.}
\label{tab:fulltable}
\end{table}


\begin{figure}
\includegraphics[width=15cm,keepaspectratio]{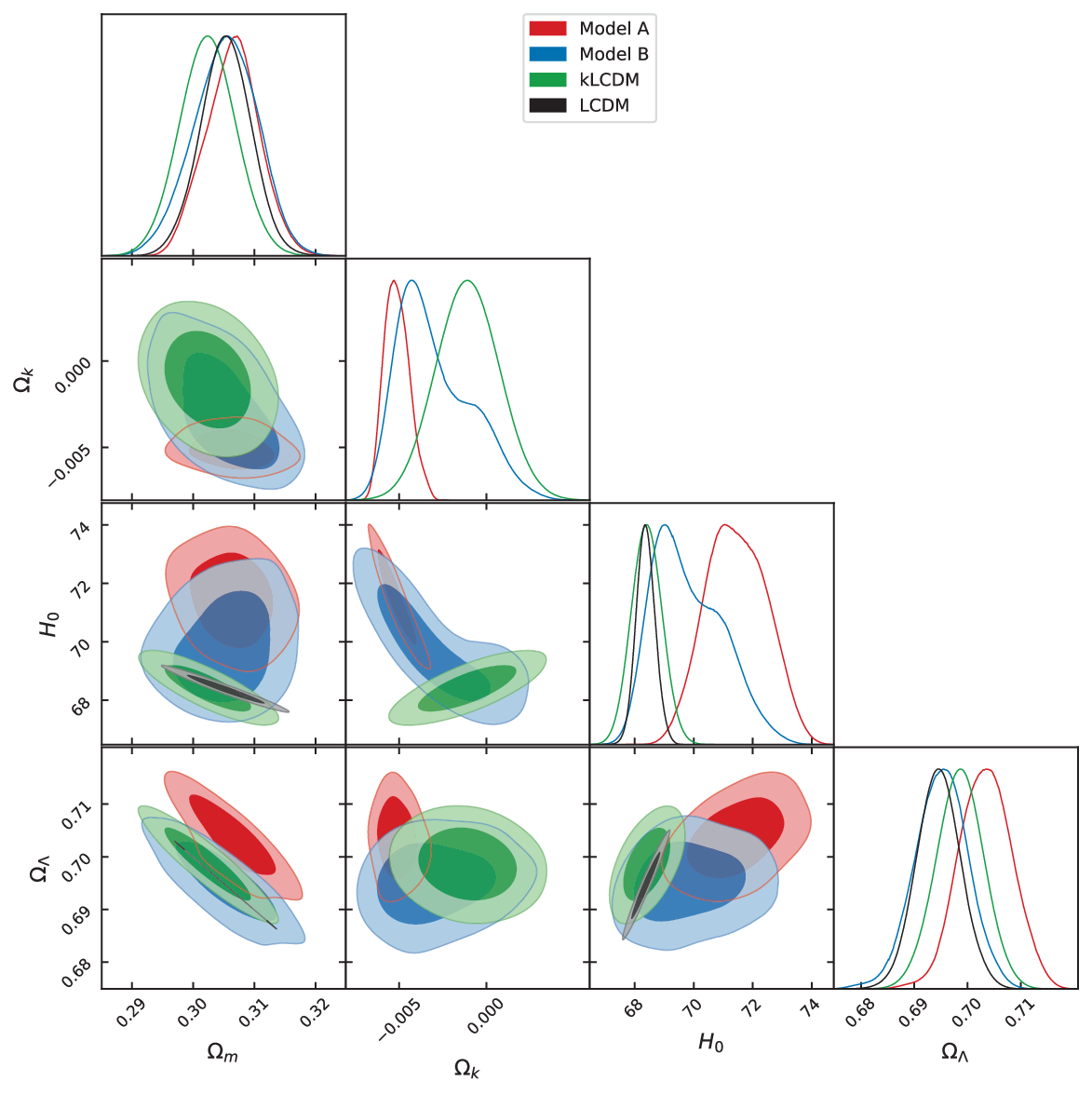}
\caption{2D joint and 1D marginalized posterior probability distributions
for $\Om_m,\Om_k,H_0$, and $\Om_\La$, obtained within our two models
{\bf A}, {\bf B}
and the standard cosmology models,
spatially-flat
LCDM,
spatially-non-flat
kLCDM.
Contour plots are shown at $1 \sigma~ (68 \%)$ and $2 \sigma~ (95 \%)$ CL.}
\label{Essential_plot}
\end{figure}

\begin{figure}
\includegraphics[width=12cm,keepaspectratio]{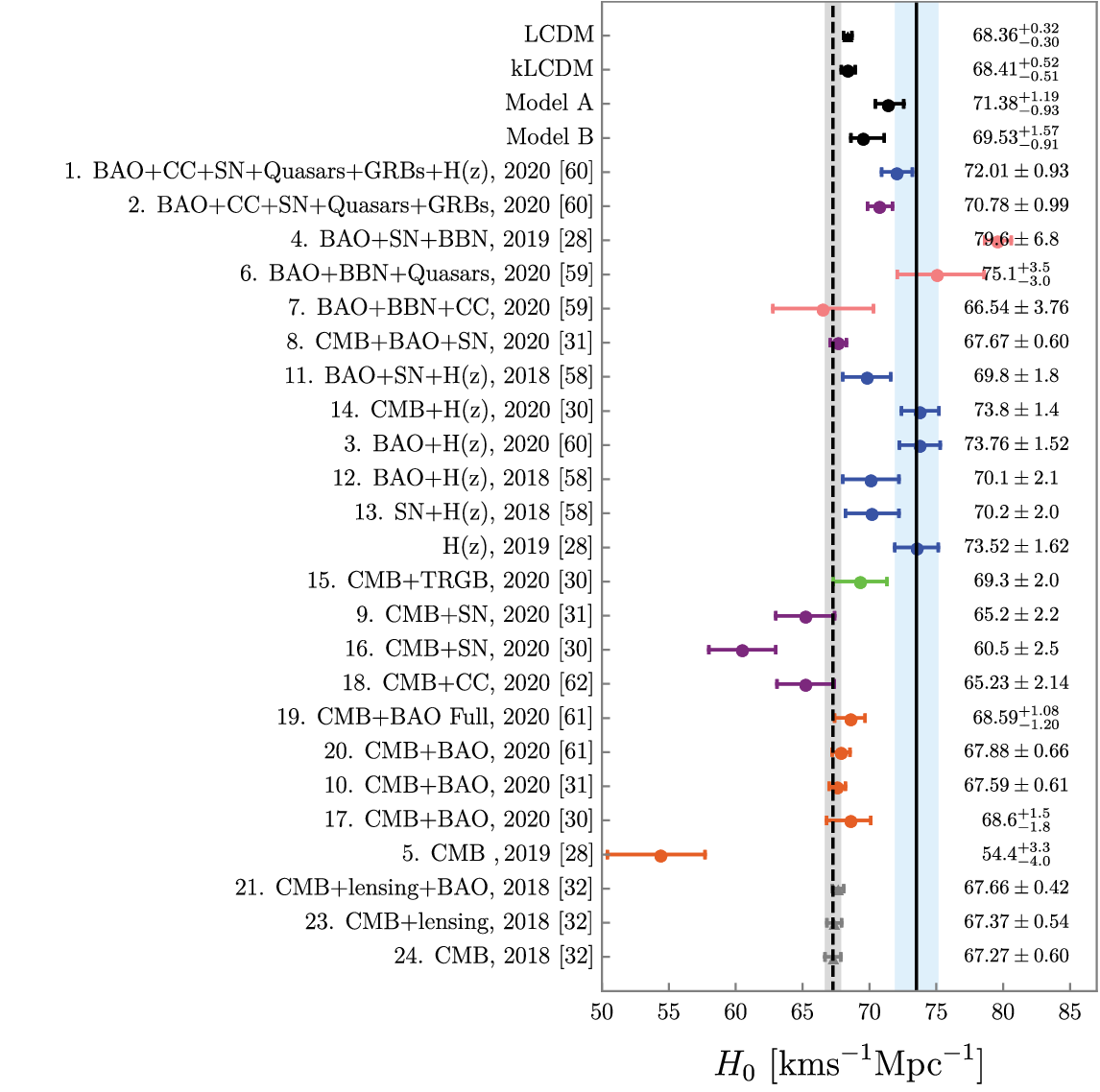}
\caption{Constraints on the Hubble constant $H_0$ at $1 \sigma$ CL
obtained by different measurements
vs. our constraints
for the models {\bf A}, {\bf B}, LCDM, and kLCDM, using all the data sets
in Table 1{. In the different measurements, we use all the {\it non-flat}
universe analysis (their companions with the same left numbers in Fig. 3) except the $H(z)$ data \cite{DiVa:2019} and the bottom three data \cite{Planck:2018}, and the full analysis
(without priors) for all the CMB data, in contrast to our analysis
with the CMB priors.} The grey vertical band corresponds the
value $H_0=67.27\pm 0.60~ {\rm km s^{-1}Mpc^{-1}}$ as reported
by Planck 2018 team \cite{Planck:2018} within a LCDM scenario.
The blue vertical band corresponds to the value
$H_0=73.52 \pm 1.62~ {\rm km s^{-1}Mpc^{-1}}$ from the recent direct
local measurement using Cepheids
\cite{Ries:2019}.  }
\label{H0_comparison}
\end{figure}

\begin{figure}
\includegraphics[width=12cm,keepaspectratio]{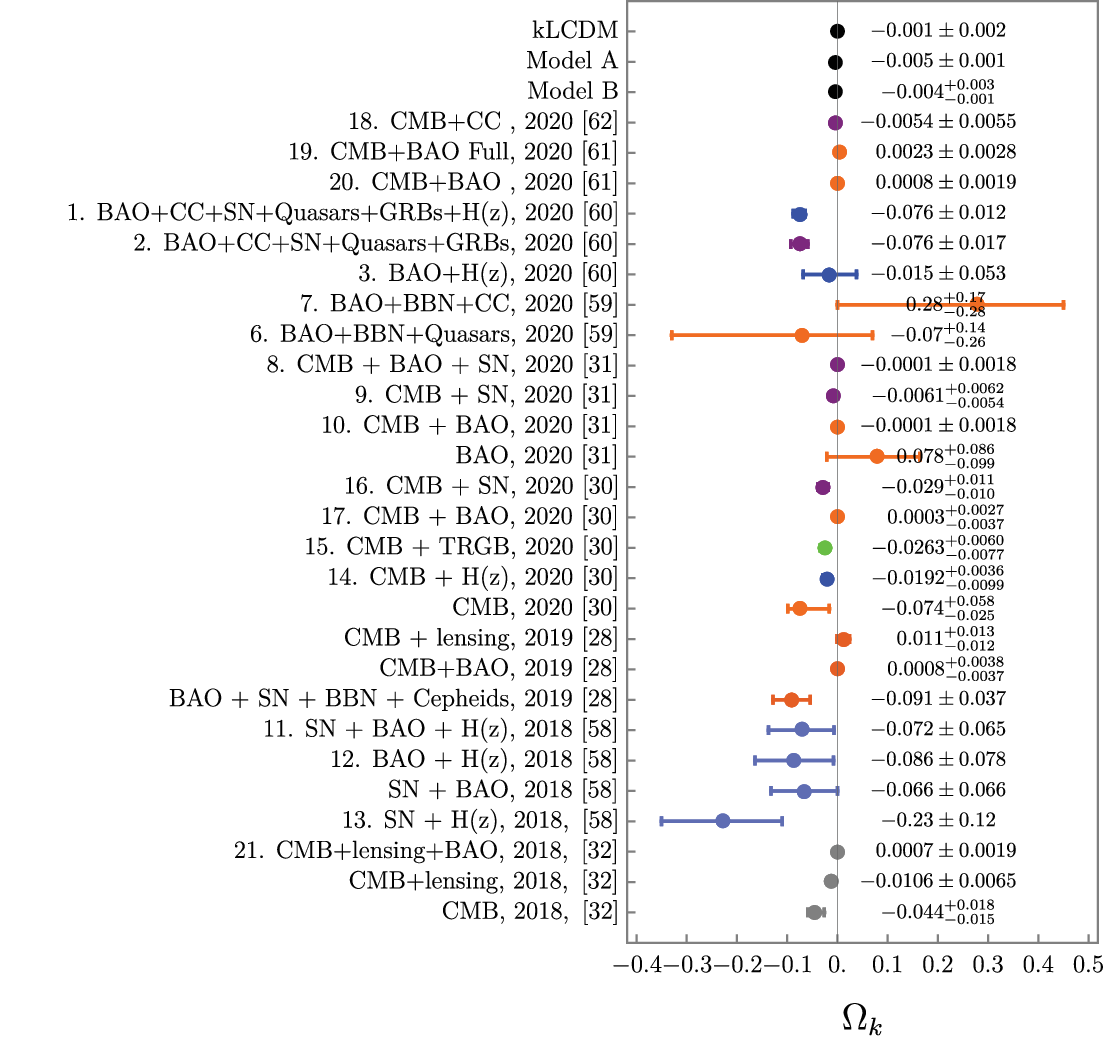}
\caption{Constraints on $\Om_k$ at $1 \sigma$ CL obtained by different measurements {based on a minimal extension to LCDM} vs. our constraints
for the models {\bf A}, {\bf B}, and kLCDM, with all the data sets
in Table 1. {The corresponding
companions in Fig. 2 have the same left numbers.}}
\label{Ok_comparison}
\end{figure}

1. For our two main models {\bf A} and {\bf B},
{constraints}
with all the data sets, including BAO, {show
some reduced tensions} in
the Hubble constant $H_0$, from
the direct local measurements
using Cepheids
 \cite{Ries:2018}
 (see Fig. 2 for a comparison with other
 previous measurements {
\cite{Park:2018,Nune:2020,Beni:2020,Vagnozzi:2020,Vagnozzi:2020dfn}}). In particular, for model {\bf A}, the tension
 is reduced to within $1 \si$ even if BAO data is included.

 {On the other hand,} for $\Om_m$ and $\Om_\La$, there are only some slight differences
 from the standard LCDM or kLCDM, and the { conventional constraint}
 of
 $\Om_m\approx 0.3$ and $\Om_\La \approx 0.7$ \cite{Ries:1998} is still
 robust even in Ho\v{r}ava
 cosmology. {This result is important because it may indicate a
 resolution of the {\it discordance} problem in the LCDM paradigm, which constrains $\Om_m \sim 0.5, \Om_\La \sim 0.5$ when a {\it closed} universe is considered as preferred by Planck \cite{DiVa:2019,Hand:2019}, but {\it incompatible} with other local measurements. However, since our result may not be its complete confirmation because we do not have the results for separate data sets but only for their combinations, as noted in the footnote No. 7.}

 {The constraint of $\Omega_r$, for model {\bf A},
 is not overlapping within $2 \sigma $ CL of the Planck 2018
 data,
 $({9.179^{+ 0.441}_{-0.431}}) \times10^{-5}$ (CMB+lensing+BAO)
 \cite{CMB Wiki}.
However, it is not surprising that $\Omega_r$, which is a derived quantity from the
{standard} formula below Eq. (13) and reduced by a factor $H_0^{-2}$ from the increased constraint on $H_0$, which is in about $4 \sigma$ tension with the Planck 2018 data $H_0=67.27\pm 0.60~ {\rm km s^{-1}Mpc^{-1}}$
\cite{Planck:2018}, has a similar
tension with the Planck data.}

{The constraints of $\Om_\om$ are rather different for models {\bf A} and {\bf B}. $\Om_\om$ is a newly introduced parameter in our models
and it has no {\it a priori} known constraint.
But our results show
that $\Om_\om<0$ is preferred
at $1.6 \si$ for model {\bf A} so that $\Om_k \Om_\om>0$, whilst poorly constrained for model {\bf B}.}
 \\

2. $\Om_k$ is peculiar in two aspects as follows. First, the result for model {\bf A}
shows a higher precision, {\it i.e.}, 
narrower MCMC contour
(see $\Om_k$ vs. $\Om_m$ in Fig. 1, for example) and, as a result, a
closed universe, {\it i.e.}, $\Om_k <0$, is more strongly preferred than in
kLCDM (cf. \cite{DiVa:2021}). (See Fig. 3 for the comparisons with other
results {based on a minimal extension to LCDM}) This is peculiar because the lower precision ({\it i.e.}, wider distribution)
is normally expected with the addition of more parameters, as can be seen in
$\Om_b h^2, \Om_r$, and $H_0$ for {\bf A} and {\bf B}, in contrast
to kLCDM.
Second, the correlation between $H_0$ and $\Om_k$ is quite different for
{\bf A} or {\bf B} and kLCDM: As $-\Om_k$ increases, {\it i.e.}, the universe
tends more towards a closed universe, $H_0$ increases for {\bf A} and {\bf B},
{\it i.e.}, the two parameters are {\it anti-correlated}, whereas the
situation is the opposite for kLCDM.
This property for $H_0$ resolves the problem of
Hubble constant tension
in kLCDM \cite{DiVa:2019,DiVa:2020} \footnote{It is interesting that a similar
effect of curvature has been noted earlier in \cite{Ries:2019} [see Fig. 4]
though it would be hardly justified in the
data sets with the CMB, based on GR \cite{Vagnozzi:2020,Vagnozzi:2020dfn}} {(see
Fig. {5} for the $H_0$ vs. $\Omega_k$ tension in the measurements of Figs. 2 and 3).} These peculiar properties seem to be due to the non-linear coupling of $\Om_k$ in the Friedmann equations (\ref{H_eq:Model A}) and (\ref{Model_B}). \\

{
3. In our analysis, we use the CMB distance priors, or shift parameters,  instead of the full CMB data (for the details, see Appendix {\bf A}). It is possible that this choice may lead to a lower statistical weight generally, though it has been widely used and is a convenient substitute of the full CMB data. However, the comparison of our analysis for the kLCDM or LCDM case and the full CMB analysis in Fig. 2 when combining other data sets, including BAO, shows that it has just a
small effect as can be seen. For example,
CMB+BAO+SN \cite{eBOS:2020},
CMB+BAO Full \cite{Vagnozzi:2020},
CMB+lensing+BAO \cite{Planck:2018}
(with or without $\Omega_k$) that use the full CMB data, show a few
percent effects ($< 2 \%$), in comparison with our LCDM, kLCDM
results with the CMB priors \footnote{There is a more systematic analysis
on the test of the CMB priors which
supports the use of the CMB priors for the background
analysis of dark-energy dynamics
\ci{Zhai:2019}. It would be an interesting
problem
to confirm the accuracy of the CMB priors
for our Ho\v{r}ava cosmology model also in a more systematic way.}.
\\}
{
4. The constraint of $\Delta N_{\rm eff}$, for model {\bf A}
(considering its central value), is {\it consistent} with SPT-3G 2018
data (alone), $
 0.65\pm 0.70$ (CMB)\footnote{In this case, the constraint of
 $H_0$ is given by $73.5 \pm 5.2~ {\rm km s^{-1}Mpc^{-1}}$. } \ci{SPT:2021}
 by $0.31 \sigma$, while in
a {\it tension} with Planck 2018 data, for example,
$2.7-2.8 \sigma$ of $
-0.06^{+0.34}_{-0.33}$ (CMB+lensing+BAO) \ci{Planck:2018}. Since, in our MCMC analysis, we have not included
SPT-3G 2018 data, which gives better constraints at the higher
angular multipoles $l>1000$, the agreement of our constraint of
$\Delta N_{\rm eff}$ might give an independent support of our model
{\bf A}.
 However, due to the lack of data with both $\Delta N_{\rm eff}$ and $\Omega_k$, which would be quite (anti-)correlated
 (see Fig. 4), as the {\it varying} parameters in SPT and Planck,
 the exact 
 quantification of (in)compatibility is not available yet.
 But we can generally expect some tensions with Planck 2018 in $\Delta N_{\rm eff}$ as in the case of $H_0$, due to the correlation of $\Delta N_{\rm eff}$ and $H_0$ in Planck 2018
 \ci{Planck:2018,SPT:2021},
 as well as in SPT-3G 2018 \ci{SPT:2021} and our case (Fig. 4).

 It is known that primordial
 gravitational waves may add to the effective relativistic degrees of freedom
 \ci{Mang:2001}.
 For example, in a recently proposed scenario linking
 $\Delta N_{\rm eff}$
 with the gravitational waves from primordial black holes \ci{Flor:2020},
 it is possible to have $\Delta N_{\rm eff(GW)}\sim 0.1-0.2$ which may increase the standard value $N_{\rm eff}=3.046$ and its associated radiation parameter $\Om_r$, slightly.
In some different scenarios, it has been also found that the primordial
gravitational waves do not solve but alleviate the $H_0$-tension problem
\ci{Grae:2018}. However, in all those cases,
a spatially-{\it flat} universe is assumed {\it implicitly} and the
effect of a non-flat universe seems to
still be an open problem. Moreover,
due to some inaccuracies of the (as adopted by us)
CMB priors for the primordial power spectrum in a non-flat universe \cite{Zhai:2019}, in this paper, we will not quantify its effect in the $H_0$-tension problem further which needs another full data analysis as well.
 \\}

5.
From
Bayes theorem, the Bayesian evidence $E({\cal D}|{\cal M})$ for a model ${\cal M}$ with the total data sets ${\cal D}$ is given by the integration over the model parameters ${\bf \theta}$
\begin{\eq}
E({\cal D}|{\cal M})= \int d{\bf \theta}~ {\cal L} ({\cal D}|{\bf \theta},{\cal M} )~
\pi ({\bf \theta}|{\cal M} ),
\end{\eq}
where ${\cal L} ({\cal D}|{\bf \theta},{\cal M})$ is the likelihood ${\cal L} ({\cal D}|{\bf \theta},{\cal M})
\equiv exp [-\chi^2 ({\cal D}|{\bf \theta},{\cal M})/2]$ in which the total
$\chi^2$ 
is obtained by summing
$\chi^2 ({\cal D}_i|{\bf \theta},{\cal M})$ over all the data sets
${\cal D}_i$. $\pi ({\bf \theta}|{\cal M})$ is the prior probability, which we
 have assumed to be flat,
{\it i.e.,} no prior information on $\bf \theta$, in order to be as agnostic
as possible:
our only priors \footnote{In general, if the prior range is too small, the parameter chain can
   be seen to `hit a wall' at one end of the prior (which serves as a hard cutoff), but we have not observed
   this behavior in our analysis.
   For the 2D contours in Fig. 4, in particular for $\Omega_k$ and $\Omega_w$,
   we zoom out enough
   to show the details of model {\bf A}
   as clearly as possible but
   at least without cutting out $2 \sigma$ CL, in compatible with
   $2 \sigma$ contours in Fig. 1.
   However, for the 1D contours of $1 \sigma$ CL in Fig. 4, we have the desired
   vanishing tails of the posterior distributions at the boundary of the
   priors,
   and all of our results have passed the convergence tests
   in \cite{Dunk:2004}.
   }
are $0<\Omega_b<\Omega_m,~ 0<H_0<100$ for LCDM, kLCDM; in addition, $\Omega_k \neq 0,~
\Delta N_{\rm eff} \neq 0$ for model {\bf A} and $\Omega_\Lambda \neq 0$ for model {\bf B}{, in order to avoid the singularity of the corresponding Friedmann equations (15) and (16), respectively \footnote{The infinitesima widths of the excluded parameter regions, which are basically discrete, around the singularities are not fixed but randomely chosen in the MCMC analysis. But, the smooth contours around the singuraities in Fig. 4 show that our chosen priors work well.}.} The differences of $\chi^2_{\rm min}$ for the models {\bf A} and {\bf B}, with respect to LCDM, $\Delta \chi^2_{\rm min}={-14.3, -7.8}$ indicate the notable improvements of fitting to the given data sets, in contrast
to the smaller difference $\Delta \chi^2_{\rm min}={-2.9}$ for kLCDM.\\

For a more quantitative comparison of the models, we consider the Bayes factor
\begin{\eq}
B_{ij} \equiv \f{E({\cal D}|{\cal M}_i)}{E({\cal D}|{\cal M}_j)},
\end{\eq}
\\
which quantifies the preference
for model ${\cal M}_i$ against
model ${\cal M}_j$, using the (revised) Jeffrey's scale \cite{Jeff:1939,Kass:1995,Ness:2012}: weak ($0 \leq {\rm ln} B_{ij} <1.1$), definite ($1.1 \leq {\rm ln} B_{ij} <3$), strong
($3 \leq {\rm ln} B_{ij} <5$), very strong ($5 \leq {\rm ln} B_{ij}$). Table 1 gives the Bayes factors for model ${\bf A}$ and ${\bf B}$ with
respect to LCDM, $\ln{B_{ij}}={+6.2,
+3.1}$,
{\it i.e.}, ``very strong" and ``strong" evidences, respectively, against the flat LCDM, in contrast to the ``definite"\footnote{This is in contrast to the ``strong" evidence for kLCDM with $\ln{B_{ij}}=+3.3$ for
CMB data alone \cite{DiVa:2019}.} evidence for kLCDM with
$\ln{B_{ij}}={+1.1}
 $
\footnote{{The detailed {\it decisiveness} of the results can depend on the adopted scales. For example, in Trotta's revisited scale \cite{Trot:2008}, our results show ``strong", ``moderate", and``weak" eveidence, respectively.} }.
 \\

{6}. The theory parameters can be written in terms of cosmological parameters as
\begin{\eq}
\label{theory_parameter}
\om=-\f{2 k \Om_\om}{R_0^2 \Om_k},~\La_W=-\f{2 k \Om_\La}{R_0^2 \Om_k},
~\mu^2=\left(\f{\Om_k }{-k} \right)\f{H_0 R_0 M_P}{\Om_{\La} L_P},
\end{\eq}
where $M_P$ and $L_P$ are the Planck mass and length, respectively, with
$M_P/L_P =c^2/8 \pi G=\al^{-1}$. Then
we obtain their best-fit values\footnote{It is interesting to note that this is
the case of $\om<0,~\om<2 \La_W$ in which the observer region is located between the
inner and outer (black hole) horizons with the {\it cutting-edge} (surface-like)
singularity of the space-time (or, the end of world) inside the inner horizon
\cite{Argu:2015}.}, $(\bar{\om},~ \bar{\La}_W, ~\bar{\mu})=(-275.94,~ 270.99,~ 0.0062), ~(173.51,~390.24,~0.0043)$, where $\bar{\om}\equiv \om R_0,~\bar{\La}_W\equiv \La_W R_0,~\bar{\mu}\equiv \left(\mu^2 L_P/H_0 R_0 M_P \right)^{1/2}$, for the models {\bf A} and {\bf B}, respectively\footnote{These correspond to the CPL parameters,
$(\om_0, ~\om_a)=(-1.005,~ -0.010), ~(-0.998,~{0.004})$
for the expansion of $\om_{\rm DE}=\om_0+\om_a (1-a)+\om_b (1-a)^2+\cdots$, near the current epoch $a=1$.}. There are large errors (see Table 2 and Fig. {6}) due to the non-linearity of the relation (\ref{theory_parameter}) but their best-fit values are distributed near what can be obtained by plugging in the obtained best-fit values of
observational parameters in Table 1,
$(\bar{\om},~ \bar{\La}_W, ~\bar{\mu})=(-299.20,~ 281.20,~ 0.00596), ~(167.50,~347.50,~0.00479)$.
These would be the
first full (cosmological) determination of the theory {parameters}
 (for the earlier determinations \footnote{There are sign errors for the results of $\bar{\om}$ and $k$ in \cite{Park:2009}.}, see \cite{Park:2009}).
\\

In conclusion, we have tested the spatially non-flat universe in
standard cosmology within \Ho~ gravity. We have obtained the ``very strong" and
``strong" Bayesian evidences against flat LCDM,
for our two models {\bf A} and {\bf B}, respectively. Moreover, the MCMC analysis shows (a)
the preference of a closed universe, (b) a reduction of the Hubble
constant tension,
{and (c) a positive result on the discordance problem}, even if BAO data is included. {It is remarkable that just the use of (14) for model {\bf A}, which gives a {\it novel} relation between $\Om_\La, \Om_k, \Om_r$ via $\De N_{\rm eff}$ which are otherwise unrelated, produces the difference in the results of the two models.}

{The reduced} Hubble tension
may be related to a natural
inclusion of the (early) dark radiation within the dynamical dark energy
and its associated contribution to $\Delta N_{\rm eff}= 0.87$ {for model {\bf A}},
{and the} peculiar contribution of the curvature $\Om_k <0$,
{as} have been previously considered on phenomenological
{ground}
\cite{Ries:2018,Knox:2019}.
However, the resolution does not seem to be quite complete yet \cite{Ries:2020}; this might be due to the fact that our observational data
{may}
not be completely model independent and are based on known physics. For example, as noted earlier, the BBN-like parametrization (\ref{BBN_DR_relation}), which has been used for model {\bf A}, is based on standard particle physics model with the phenomenological parameter $\Delta N_{\rm eff}$. Our MCMC analysis implies that the phenomenological approach of model {\bf A} is a good approximation of the BBN, constraining $\Om_k$ more precisely and reducing the Hubble constant tension, {\it i.e.}, increasing $H_0$, with the almost doubled Bayesian evidence compared to model {\bf B}. It would be a challenging problem to thoroughly consider the effect of anisotropic scaling beyond the standard model BBN and revisit the Hubble tension problem to see whether a complete resolution can be found or not. Finally, the analysis of the cosmological
perturbations {for the non-flat universe and with the full use of Boltzmann solvers such as CAMB/Class {\cite{Lewi:1999,Lesg:2011}}} would also be an important arena for studying standard
cosmology, like the cosmic-shear or $\si_8$ tension \cite{DiVa:2019}.


\section*{Acknowledgments}
We would like to thank Wayne Hu, Hyung-Won Lee, Seokcheon Lee, Seshadri Nadathur, Chan-Gyung Park, Vincenzo Salzano, and
Eleonora Di Valentino for helpful discussions. {We also thank to an anonymous referee for helpful comments which have inspired us to improve our paper.}
{NAN and}
MIP were supported by Basic Science Research Program through the National Research Foundation of Korea (NRF) funded by the Ministry of Education, Science and Technology {{(2020R1A2C1010372) [NAN]}, (2020R1A2C1010372, 2020R1A6A1A03047877) [MIP]}.

\appendix{\label{app_A}}

\begin{section}
{Cosmological Data Sets and $\chi^2$ Measures}
\end{section}

In this Appendix, we present some more details on the cosmological data sets
and their $\chi^2$ measures as used for the statistical analysis.
\\

{\it 1. Cosmic Microwave Background (CMB)}\\

The CMB data is given by the shift
parameters which describe the location of the first peak in the
temperature angular power spectrum. We use the shift parameters
for Planck 2018 data \cite{Zhai:2018}, {which includes temperature
and polarization data, as well as CMB lensing, and} given by
the ratio between the model being tested and the LCDM model ($\mathbf{x}$ is a vector containing the
model parameters) with
a canonical cold dark matter 
as \cite{Wang:2015}
\begin{align}
   \nonumber R(\mathbf{x}) &=100 \sqrt{\Omega_mh^2}~
   \frac{d^c_A (z_*,\mathbf{x})}{c}, \\
    \nonumber \ell_a(\mathbf{x})& = \pi
    \frac{d^c_A (z_*,\mathbf{x})}{r_s(z_*,\mathbf{x})},
\end{align}
as well as $\Omega_b h^2$,
with the reduced Hubble constant $h \equiv H_0/100~ {\rm km s^{-1} Mpc^{-1}}$. Here, $d^c_A (z_*,\mathbf{x})$ and $r_s(z_*,\mathbf{x})$ are the comoving
angular-diameter distance and the sound horizon
\begin{\eq}
   d^c_A (z,\mathbf{x}) &=&\f{c}{H_0} \f{1}{\sqrt{\Om_k}}~ {\rm Sinh} \left[ \sqrt{\Om_k}  \int_0^z \frac{dz'}{E (z',\mathbf{x})} \right], \\
   r_s (z,\mathbf{x}) &=&\int^\infty_z \frac{c_s(z') dz'}{H (z',\mathbf{x})},
\end{\eq}
respectively, where $E(z,\mathbf{x}) \equiv H(z,\mathbf{x})/H_0$ and $c_s$ is the sound speed
\begin{equation}
    c_s(z) = \f{c}{\sqrt{3[1+R_b (1+z)^{-1}]}}
\end{equation}
and
\begin{equation}
    R_b = 31500~ \Omega_b h^2 \left (T_{\rm CMB}/2.72 \right)^{-4}
\end{equation}
at the photon-decoupling redshift $z_*$,
given by \cite{Hu:1995}
\begin{align}
    \nonumber z_* &= 1048\left[1+0.00124~(\Omega_bh^2)^{-0.738}\right] \left[1+g_1~(\Omega_mh^2)^{g_2}\right], \\
    \nonumber g_1 &= 0.0783~(\Omega_bh^2)^{-0.238}\left[1+39.5~(\Omega_bh^2)^{-0.763}\right]^{-1}, \\
    g_2 &= 0.560~ \left[1+21.1~(\Omega_bh^2)^{1.81}\right]^{-1}.
\end{align}

Then, we obtain the $\chi^2$ measure for Planck 2018 as
\begin{equation}
    \chi^2_{\rm Planck} = \left(\Delta\mathcal{F}_{\rm Planck}\right)^{\rm T} 
    C^{-1}_{\rm Planck} 
    \Delta\mathcal{F}_{\rm Planck},
\end{equation}
where $\Delta\mathcal{F}_{\rm Planck}\equiv \mathcal{F}_{\rm  Planck (th)}-\mathcal{F}_{\rm Planck (obs)}$ is the difference between the theoretical and observed distance modulus for a vector formed from the three shift parameters $\mathcal{F}_{\rm Planck}\equiv \left\{ R, \ell_a, {\Om_b} h^2 \right\}$ and $C^{-1}_{\rm Planck}$ is the inverse covariance matrix \cite{Wang:2015}.\\

{\it 2. Baryon Acoustic Oscillations (BAO)}\\

The BAO consists of several data sets:\\

{\bf SDSS-BOSS:}
This includes the data points from the Sloan Digital Sky Survey III - Baryon Acoustic Oscillation Spectroscopic Survey (SDSS-BOSS), DR12 release \cite{BOSS:2016}, with associated redshifts $z_B=\{0.38, 0.51, 0.61\}$.
The pertinent quantities for the BOSS data are
\begin{align}
    &\nonumber d^c_A (z,\mathbf{x}) ~ \f{r_s^{\rm fid}(z_d)}{r_s(z_d,\mathbf{x})},\\
    &H(z,\mathbf{x})~ \f{r_s(z_d,\mathbf{x})}{r_s^{\rm fid}(z_d)}
\end{align}
at the {\it dragging} redshift $z_d$ which can be approximated as \cite{Eise:1997}
\begin{align}
    \nonumber z_d &= \frac{1291~ (\Omega_mh^2)^{0.251}}{1+0.659~ (\Omega_mh^2)^{0.828}}
    \left[1+b_1~ (\Omega_bh^2)^{b_2}\right],\\
    \nonumber b_1 &= 0.313 ~ (\Omega_mh^2)^{-0.419}\left[1+0.607 ~ (\Omega_mh^2)^{0.6748}\right],\\
    b_2 & = 0.238 ~(\Omega_mh^2)^{0.223}.
\end{align}
Here, $r_s^{\rm fid}(z_d)$ is the same quantity evaluated for a fiducial/reference model and we take $r_s^{\rm fid}(z_d)=147.78$ Mpc \cite{BOSS:2016}.
We then obtain its $\chi^2$ measure as
\begin{\eq}
\chi^2_{\rm BOSS} = \left(\Delta\mathcal{F}_{\rm BOSS}\right)^{\rm T} 
C^{-1}_{\rm BOSS} 
\Delta\mathcal{F}_{\rm BOSS}
\end{\eq}
with $\Delta\mathcal{F}_{\rm BOSS} \equiv \mathcal{F}_{\rm BOSS (th)}-\mathcal{F}_{\rm BOSS (obs)}$ for a vector
$\mathcal{F}_{\rm BOSS} \equiv \left\{d^c_A (z_B) r_s^{\rm fid}(z_d)/r_s(z_d),~ H (z_B) r_s(z_d)/r_s^{\rm fid}(z_d)\right\}$ and
$C^{-1}_{\rm BOSS}$ is the inverse covariance matrix found in \cite{BOSS:2016}.\\

\textbf{SDSS-eBOSS:}
There is one more data point in the extended Baryon Acoustic Oscillation Spectroscopic Survey (eBOSS) \cite{Ata:2017} at $z = 1.52$, which gives the value
\begin{equation}
    D_V(1.52,\mathbf{x}) ~\frac{r_s^{\rm fid}(z_d)}{r_s(z_d,\mathbf{x})} = 3843 \pm 147 ,
\end{equation}
where $D_V$ is defined as
\begin{equation}
    D_V(z,\mathbf{x}) = \left[d^c_A (z,\mathbf{x})^2 \frac{cz}{H(z,\mathbf{x})}\right]^{1/3}.
    \label{D_V}
\end{equation}
We have then its $\chi^2$ measure as
\begin{equation}
    \chi^2_{\rm eBOSS} = \frac{\Delta\mathcal{F}_{\rm eBOSS}}{\sigma^2_{\rm eBOSS}}
\end{equation}
with $\Delta\mathcal{F}_{\rm eBOSS}\equiv \mathcal{F}_{\rm eBOSS (th)}-\mathcal{F}_{\rm eBOSS (obs)}$ for
$\mathcal{F}_{\rm eBOSS}\equiv D_V(z=1.52) r_s^{\rm fid}(z_d)/r_s(z_d)$ and the measurement error $\sigma_{\rm eBOSS}$.
\\

\textbf{SDSS-BOSS-Lyman-$\alpha$:}
The Quasar-Lyman-$\alpha$ forest from SDSS-III BOSS RD11 \cite{deSa:2019} gives the two data points as
\begin{align}
    \nonumber & d^c_A (z=2.34,\mathbf{x})/r_s(z_d,\mathbf{x}) = 36.98^{+1.26}_{-1.18},\\
    &c/H(z=2.34,\mathbf{x}) r_s(z=z_d,\mathbf{x}) = 9.00 \pm 0.22,
\end{align}
and we obtain the corresponding $\chi^2$ measure as
\begin{\eq}
\chi^2_{\rm Lyman-\al}= \left(\Delta\mathcal{F}_{\rm Lyman-\al}\right)^{\rm T} 
C^{-1}_{\rm Lyman-\al} 
\Delta\mathcal{F}_{\rm Lyman-\al}
\end{\eq}
with $\Delta\mathcal{F}_{\rm Lyman-\al}\equiv \mathcal{F}_{\rm Lyman-\al (th)}-\mathcal{F}_{\rm Lyman-\al (obs)}$ and
$\mathcal{F}_{\rm Lyman-\al}\equiv \{ d^c_A (z=2.34)/r_s(z_d),~ c/H(z=2.34) r_s(z_d)\}$.\\

\textbf{WiggleZ:}
This includes the data from the WiggleZ Dark Energy Survey at redshift points $z_{\rm W}=\{0.44,0.6,0.73\}$ \cite{Blak:2012}. Here, the pertinent quantities are the acoustic parameter
\begin{equation}
    A(z,\mathbf{x}) = 100\sqrt{\Omega_m h^2}~\frac{D_V(z,\mathbf{x})}{cz}
\end{equation}
and Alcock-Paczynski parameter
\begin{equation}
    F(z,\mathbf{x}) =\frac{d^c_A (z,\mathbf{x})H(z,\mathbf{x})}{c},
\end{equation}
where $D_V$ is defined as above (\ref{D_V}). We then obtain its $\chi^2$ measure as
\begin{\eq}
\chi^2_{\rm WiggleZ}= \left(\Delta\mathcal{F}_{\rm WiggleZ}\right)^{\rm T} 
 C^{-1}_{\rm WiggleZ} 
 \Delta\mathcal{F}_{\rm WiggleZ}
\end{\eq}
with $\Delta\mathcal{F}_{\rm WiggleZ}\equiv \mathcal{F}_{\rm WiggleZ (th)}-\mathcal{F}_{\rm WiggleZ (obs)}$ and
$\mathcal{F}_{\rm WiggleZ}\equiv \{ A (z_W),~ F (z_W) \}$.
\\

{\it 3. Type Ia Supernovae (SNe Ia)}\\

We use the recent Pantheon catalogue of Type Ia supernovae (SNe Ia) which consists of 1048 objects in the redshift range $0.01 < z < 2.26$ \cite{Scol:2017}. The data is expressed as the distance modulus
\begin{equation}
\label{mu}
    \mu(z,\mathbf{x}) = 5\log{d_L(z,\mathbf{x})} + \mu_0,
\end{equation}
where $\mu_0$ is a nuisance parameter containing
the supernova absolute magnitude and
$d_L$ is the luminosity distance
\begin{equation}
    d_L(z,\mathbf{x}) \equiv (1+z)\cdot d^c_A (z,\mathbf{x}).
\end{equation}

Removing the nuisance parameter dependence via marginalizing over
$\mu_0$ \cite{SNLS:2011}, we obtain its $\chi^2$ measure as
\begin{equation}
\label{chi^2_SN}
    \chi^2_{\rm Pantheon} =a + \log \frac{e}{2\pi} -\f{b^2}{e},
\end{equation}
where $a \equiv \Delta\mu^{\rm T} 
C^{-1}_{\rm Pantheon} 
\Delta\mu,
~ b \equiv \Delta\mu^{\rm T} 
C^{-1}_{\rm Pantheon} \cdot
{\bf 1},~ c \equiv {\bf 1}^{\rm T} \cdot
C^{-1}_{\rm Pantheon} \cdot
{\bf 1}
$
with $\Delta\mu = \mu_{\rm th}-\mu_{\rm obs}$
for each object and the inverse covariance matrix $C^{-1}_{\rm Pantheon}$ for the whole sample. Here, the theoretical value of the distance module (\ref{mu}) is given by
\begin{\eq}
\label{mu_th_SN}
  \mu_{\rm th} = 5\log \left[(1+z_{\rm hel}) \cdot d^c_A (z_{\rm cmb},\mathbf{x}) \right] + \mu_0,
\end{\eq}
where $z_{\rm cmb}$ is the CMB restframe redshift and $z_{\rm hel}$ is the {\it heliocentric} redshifts which includes the effect of the peculiar velocity \cite{Hui:2005,Wang:2013}.\\

{\it 4. Gamma-Ray Bursts (GRBs)}\\

We use the set of 79 Gamma-Ray Bursts (GRBs) in the range $1.44 < z < 8.1$, called
the Mayflower sample \cite{Liu:2014}, which was calibrated in a model independent
manner. The pertinent quantity for GRBs is the distance modulus $\mu$ and
therefore its $\chi^2$ measure is in analogue with SNe Ia in the formula (\ref{chi^2_SN}). But
the important difference is that there is no distinction between $z_{\rm cmb}$
and $z_{\rm hel}$ for the theoretical value of the distance module $\mu_{\rm th}$
in (\ref{mu_th_SN}).\\

{\it 5. Lensed Quasars}\\

We use the 6 lensed quasars from the recent release by the H0liCOW collaboration \cite{Suyu:2016,Wong:2019}. These quasars have multiple-lensed images from which
a time delay due to the different light paths can be obtained.
The time delay can be expressed as
\begin{equation}
    t(\theta,\beta) = \frac{1+z_L}{c}\frac{D_L D_{S}}{D_{LS}}\left[\frac{1}{2}(\theta-\beta)^2-\psi(\theta)\right],
\end{equation}
where
$z_L$ is the lens redshift, $\psi$ is the lensing potential, and $\theta$, $\beta$ are the angular position of the image and the source, respectively. The quantities $D_L$, $D_S$, and $D_{LS}$ are the angular-diameter distances for lens $\to$ observer, source $\to$ observer, and
source $\to$ lens, respectively,
defined as
\begin{align}
&d_A (z,\mathbf{x})=\f{d^c_A (z,\mathbf{x})}{1+z} , \\
    \nonumber & D_L = d_A (z_L,\mathbf{x}),~ \, D_S = d_A (z_S,\mathbf{x}),\\
    & D_{LS} = \frac{1}{1+z_S}\left[ (1+z_S)D_S-(1+z_L)D_L \right],
\end{align}
where $z_L$ is the source redshift.
From the time-delay distance, the combination constrained by H0liCOW, defined as  $D_{\Delta t} = (1+z_L)D_L D_S/D_{LS}$, we obtain its $\chi^2$ measure as
\begin{equation}
    \chi^2_{\rm H0liCOW} = \sum_{i=1}^{6}\frac{\left(D_{\Delta t, i}(\mathbf{x})-D_{\Delta t, i}^{\rm obs}\right)^2}{\sigma_{D_{\Delta t, i}}^2}
\end{equation}
with the observed values $D_{\Delta t, i}^{\rm obs}$ and the measurement errors $\sigma_{D_{\Delta t, i}}$.
\\

{\it 6. Cosmic Chronometers (CC)}\\

We use the early elliptical and lenticular galaxies at different redshifts
whose spectral properties can be traced with cosmic time $t$ so that they can be
used as {\it Cosmic Chronometers} (CC) by measuring the Hubble parameter
$H(z)\equiv dz/dt(1+z)$, independently on cosmological models \cite{Jime:2001}.
Then, using 25 measurements from the data set in the range $0.07 < z < 1.965$ \cite{More:2015} \footnote{{In this paper, we have not used the latest data set in \cite{More:2016} (the range $0.3<z<0.5$), due to some uncertainty in the Hubble parameter from different stellar population models.}},
we obtain its $\chi^2$ measure as
\begin{equation}
    \chi^2_{\rm CC} = \sum_{i=1}^{25}\frac{\left[H(z_i,\mathbf{x})-H_{\rm obs}(z_i)\right]^2}{\sigma_{\rm CC}^2(z_i)},
\end{equation}
where $H_{\rm obs}(z_i)$ are the measured values of the Hubble parameter and $\sigma_{\rm CC}(z_i)$ are their measurement errors.


\begin{section}
{More Details of Constraints on Parameters}
\end{section}

\begin{figure}
\includegraphics[width=17cm,keepaspectratio]{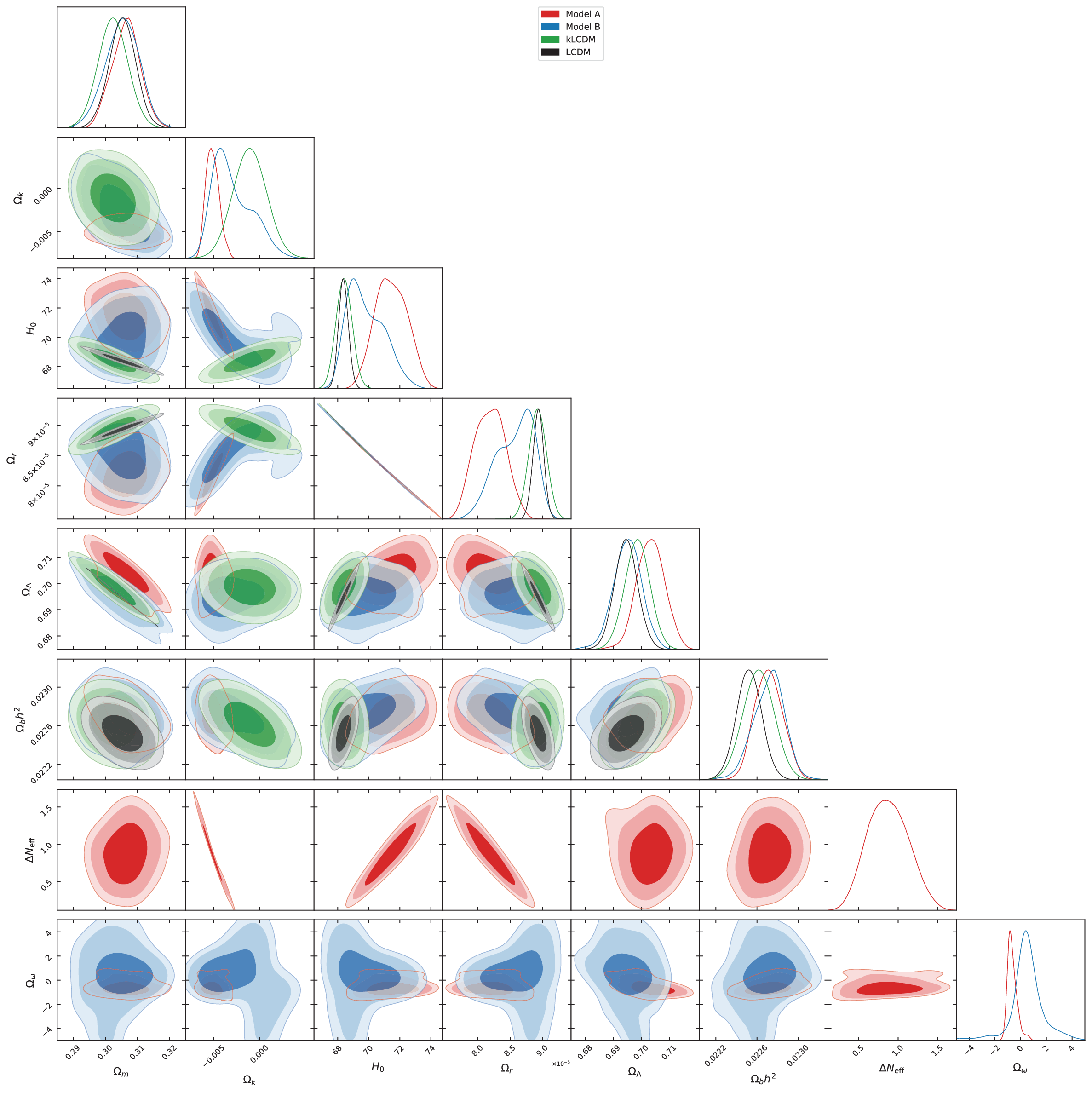}
\caption{2D joint and 1D marginalized posterior probability distributions
for all the parameters in Table 1, obtained within the models {\bf A}, {\bf B}, LCDM,
kLCDM. Contour plots are shown up to $3 \sigma~ (99.7 \%)$ CL.}
\label{Ok_comparison}
\end{figure}

\begin{figure}
\includegraphics[width=8cm,keepaspectratio]{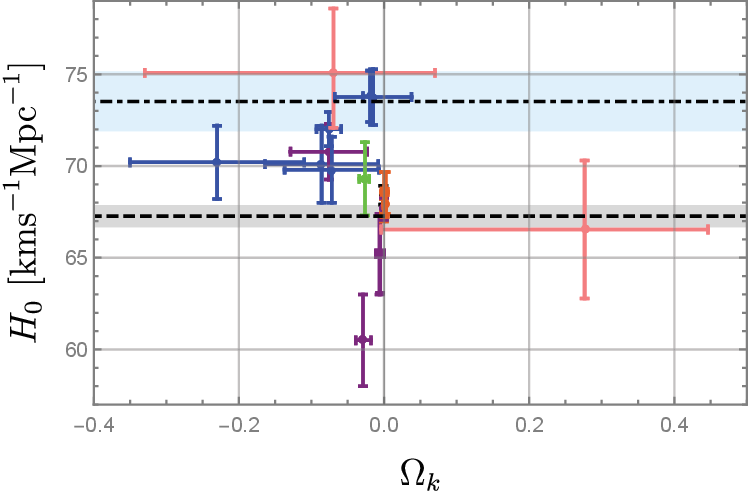}
\includegraphics[width=8cm,keepaspectratio]{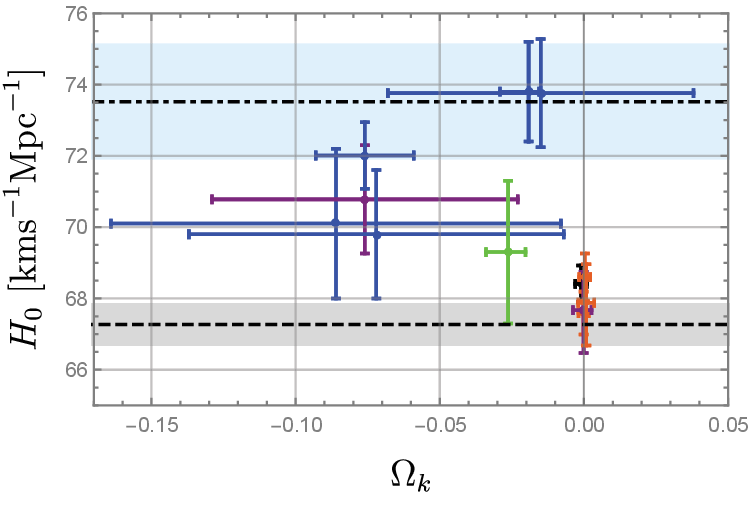}
\includegraphics[width=8cm,keepaspectratio]{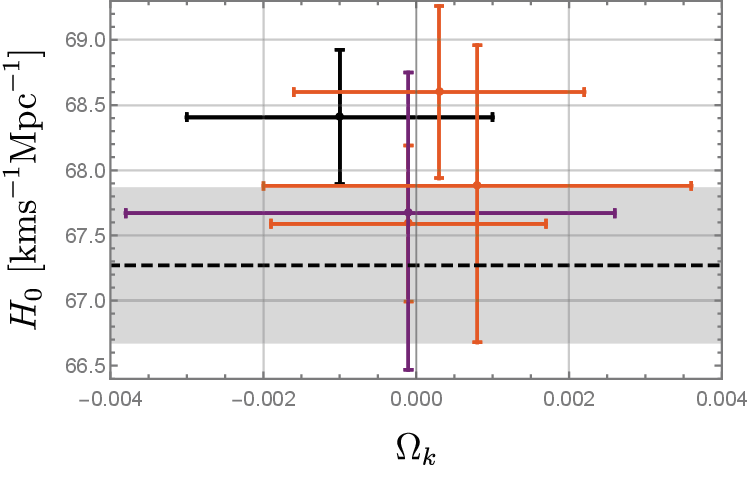}
\caption{Combined plots of the constraints on $H_0$ vs. $\Omega_k$ for the measurements in Fig. 2 and Fig. 3, based on (k)LCDM scenario. The  companion data in Figs. 2 and 3 has the same color. The grey and blue vertical bands correspond the
value of Planck 2018, $H_0=67.27\pm 0.60~ {\rm km s^{-1}Mpc^{-1}}$
\cite{Planck:2018},
and the local measurement using Cepheids,
$H_0=73.52 \pm 1.62~ {\rm km s^{-1}Mpc^{-1}}$
\cite{Ries:2019}, respectively, as in Figs. 2 and 3.  By excluding some anomalous cases (top and rightmost oranges; leftmost blue; three purple data (bottom three)) with large errors in the full data (top left) and zooming the interested region (top right), one can see a {\it rough} tendency
$H_0 \propto
-|\Omega_k |$, as in the full contour plots in \cite{DiVa:2019} [Fig. 8] or \cite{DiVa:2020} [Fig. 3].
A few data points near $\Omega_k=0$ do not show the tendency clearly but a further zooming (bottom) seems to show
another tendency of the CMB+BAO cases in
\cite{DiVa:2020} [Fig. 3].
}
\label{H0_vc_Ok}
\end{figure}

\renewcommand{\arraystretch}{1.0}
\begin{table}[h!]
\begin{tabular}{|l|c|c|}
\hline
Theory Parameters &Model {\bf A} & Model {\bf B} \\
\hline
$\bar{\omega}$ & $-275.94^{+171.38}_{-133.08}$ & $173.51^{+1357.96}_{-159.84}$ \\
$\bar{\Lambda}_W$ & $270.99^{+49.89}_{-31.65}$ & $390.24^{+1224.34}_{-117.17}$ \\
$\bar{\mu}$ & ${0.0062^{+0.0008}_{-0.0010}}$ & $0.0043^{+0.0018}_{-0.0033}$ \\
\hline
\end{tabular}
\caption{Constraints at $1 \sigma$ CL on the theory parameters
for our two cosmology models
{\bf A}, {\bf B}.}
\label{tab:fulltable}
\end{table}

\begin{figure}
\includegraphics[width=8cm,keepaspectratio]{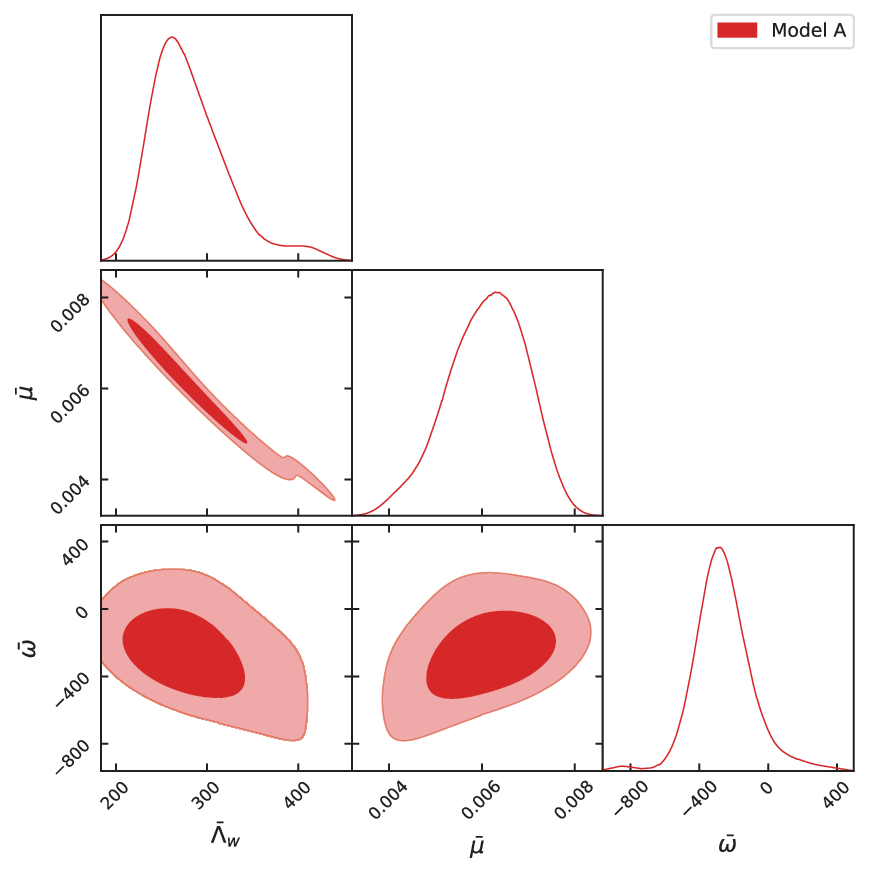}
\includegraphics[width=8cm,keepaspectratio]{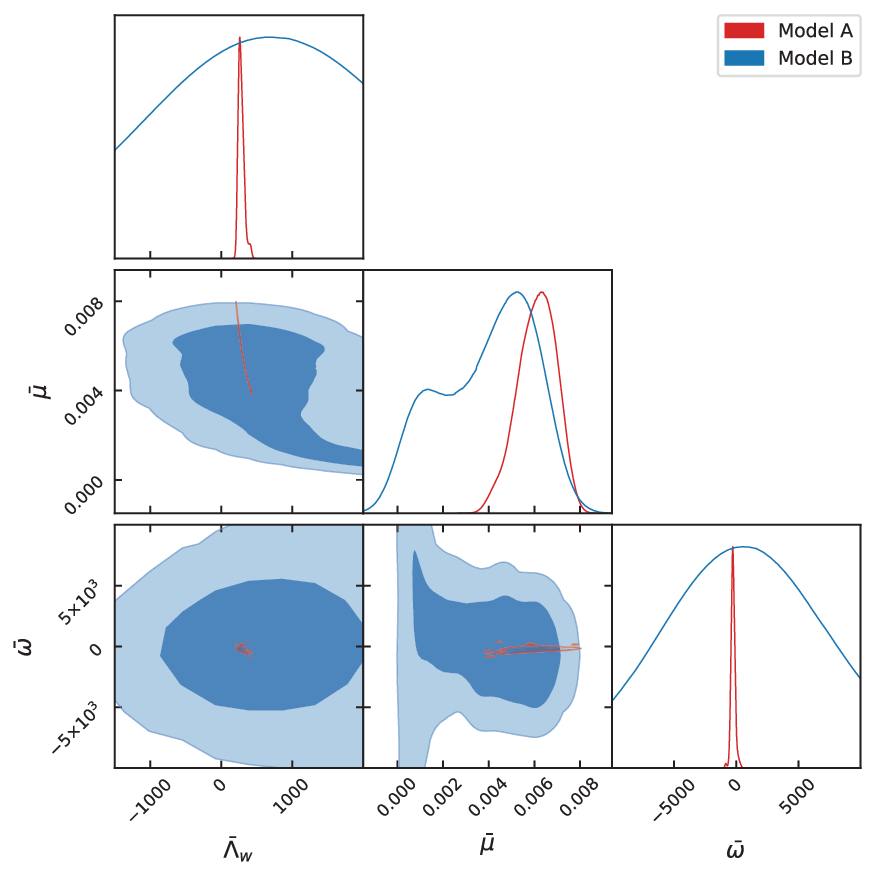}
\caption{2D joint and 1D marginalized posterior probability distributions
for the theory parameters in Table 2, obtained within the models {\bf A} (left), {\bf B} (right), up to $2 \sigma$ CL.}
\label{parameter_A}
\end{figure}

\newcommand{\J}[4]{#1 {\bf #2} #3 (#4)}
\newcommand{\andJ}[3]{{\bf #1} (#2) #3}
\newcommand{\AP}{Ann. Phys. (N.Y.)}
\newcommand{\MPL}{Mod. Phys. Lett.}
\newcommand{\NP}{Nucl. Phys.}
\newcommand{\PL}{Phys. Lett.}
\newcommand{\PR}{Phys. Rev. D}
\newcommand{\PRL}{Phys. Rev. Lett.}
\newcommand{\PTP}{Prog. Theor. Phys.}
\newcommand{\hep}[1]{ hep-th/{#1}}
\newcommand{\hepp}[1]{ hep-ph/{#1}}
\newcommand{\hepg}[1]{ gr-qc/{#1}}
\newcommand{\bi}{ \bibitem}


\begin{thebibliography}{99}

\bibitem{Frie:1922}
A.~Friedmann,
Z. Phys. \textbf{10}, 377
(1922).

\bibitem{Lema:1927}
G.~Lemaitre,
Annales Soc. Sci. Bruxelles A \textbf{47}, 49
(1927).

\bibitem{Ries:1998}
A.~G.~Riess \textit{et al.} [Supernova Search Team],
Astron. J. \textbf{116}, 1009
(1998)
[arXiv:astro-ph/9805201 [astro-ph]].

\bibitem{WMAP:2010}
E.~Komatsu \textit{et al.} [WMAP],
Astrophys. J. Suppl. \textbf{192}, 18 (2011)
[arXiv:1001.4538 [astro-ph.CO]].

\bibitem{Ries:2018}
A.~G.~Riess \textit{et al.},
Astrophys. J. \textbf{855},
136 (2018)
[arXiv:1801.01120 [astro-ph.SR]].

\bibitem{Ries:2019}
A.~G.~Riess \textit{et al.},
Astrophys. J. \textbf{876},
85 (2019)
[arXiv:1903.07603 [astro-ph.CO]].

\bibitem{Asga:2019}
M.~Asgari \textit{et al.},
Astron. Astrophys. \textbf{634}, A127 (2020)
[arXiv:1910.05336 [astro-ph.CO]].

\bibitem{DiVa:2021}
E.~Di Valentino \textit{et al.},
arXiv:2103.01183 [astro-ph.CO].

\bibitem{Peri:2021}
L.~Perivolaropoulos and F.~Skara,
arXiv:2105.05208 [astro-ph.CO].

\bibitem{Scho:2021}
N.~Sch\"oneberg \textit{et al.},
arXiv:2107.10291 [astro-ph.CO].

\bibitem{Stelle:1976}
K.~S.~Stelle,
Phys. Rev. D \textbf{16}, 953
(1977).

\bibitem{Lifs:1941} E. M. Lifshitz, Zh. Eksp. Teor. Fiz., {\bf 11}, 255 $\&$ 269 (1941).

\bibitem{DeWi:1967}
  B.~S.~DeWitt,
  Phys.\ Rev.\  {\bf 160}, 1113 (1967).

\bibitem{Hora:2009}
  P.~Ho\v{r}ava,
  Phys.\ Rev.\  D {\bf 79}, 084008 (2009)
  [arXiv:0901.3775 [hep-th]].

\bibitem{Wang:2017}
A.~Wang,
Int. J. Mod. Phys. D \textbf{26},
1730014 (2017)
[arXiv:1701.06087 [gr-qc]].

\bibitem{Deve:2021}
D.~O.~Devecioglu and M.~I.~Park,
[arXiv:2112.00576 [hep-th]].

\bibitem{Olmo:2019}
G.~J.~Olmo, D.~Rubiera-Garcia and A.~Wojnar,
Phys. Rept. \textbf{876}, 1
(2020)
[arXiv:1912.05202 [gr-qc]].

\bibitem{LIGO:2021}
R.~Abbott \textit{et al.} [LIGO Scientific, VIRGO and KAGRA],
[arXiv:2112.06861 [gr-qc]].

\bibitem{Kim:2018}
K.~Kim \textit{et al.}
Phys. Rev. D \textbf{103},
044052 (2021)
[arXiv:1810.07497 [gr-qc]].

\bibitem{Liu:2010}
M.~Liu \textit{et al.}
Gen. Rel. Grav. \textbf{43}, 1401
(2011)
[arXiv:1010.6149 [gr-qc]].

\bibitem{Li:2021}
G.~P.~Li and K.~J.~He,
JCAP \textbf{06}, 037 (2021)
[arXiv:2105.08521 [gr-qc]].

\bibitem{Emir:2017}
A.~Emir G\"umr\"uk\c{c}\"uo\u{g}lu, M.~Saravani and T.~P.~Sotiriou,
Phys. Rev. D \textbf{97},
024032 (2018)
[arXiv:1711.08845 [gr-qc]].

\bibitem{Gong:2018}
Y.~Gong \textit{et al.}
Phys. Rev. D \textbf{98},
104017 (2018)
[arXiv:1808.00632 [gr-qc]].

\bibitem{Khod:2020}
M.~Khodadi and E.~N.~Saridakis,
Phys. Dark Univ. \textbf{32}, 100835 (2021)
[arXiv:2012.05186 [gr-qc]].

\bibitem{Gupt:2021}
T.~Gupta \textit{et al.}
Class. Quant. Grav. \textbf{38},
195003 (2021)
[arXiv:2104.04596 [gr-qc]].

\bibitem{Frus:2015}
N.~Frusciante \textit{et al.}
Phys. Dark Univ. \textbf{13}, 7
(2016)
[arXiv:1508.01787 [astro-ph.CO]].

\bibitem{Frus:2020}
N.~Frusciante and M.~Benetti,
Phys. Rev. D \textbf{103},
104060 (2021)
[arXiv:2005.14705 [astro-ph.CO]].


\bibitem{DiVa:2019}
E.~Di Valentino, A.~Melchiorri and J.~Silk,
Nature Astron. \textbf{4},
196
(2019)
[arXiv:1911.02087 [astro-ph.CO]].

\bibitem{Hand:2019}
W.~Handley,
Phys. Rev. D \textbf{103},
L041301 (2021)
[arXiv:1908.09139 [astro-ph.CO]].

\bibitem{DiVa:2020}
E.~Di Valentino, A.~Melchiorri and J.~Silk,
Astrophys. J. Lett. \textbf{908},
L9 (2021)
[arXiv:2003.04935 [astro-ph.CO]].

\bibitem{eBOS:2020}
S.~Alam \textit{et al.} [eBOSS],
Phys. Rev. D \textbf{103},
083533 (2021)
[arXiv:2007.08991 [astro-ph.CO]].


\bibitem{Planck:2018}
N.~Aghanim \textit{et al.} [Planck],
Astron. Astrophys. \textbf{641}, A6 (2020)
[erratum: Astron. Astrophys. \textbf{652}, C4 (2021)]
[arXiv:1807.06209 [astro-ph.CO]].

\bibitem{Arno:1962}
R.~L.~Arnowitt, S.~Deser and C.~W.~Misner,
Gen. Rel. Grav. \textbf{40}, 1997
(2008)
[arXiv:gr-qc/0405109 [gr-qc]].

\bibitem{Muko:2009}
S.~Mukohyama,
JCAP \textbf{06}, 001 (2009)
[arXiv:0904.2190 [hep-th]].

\bibitem{Gesh:2011}
G.~Geshnizjani, W.~H.~Kinney and A.~Moradinezhad Dizgah,
JCAP \textbf{11}, 049 (2011)
[arXiv:1107.1241 [astro-ph.CO]].

\bibitem{Shin:2017}
S.~Shin and M.~I.~Park,
JCAP \textbf{12}, 033 (2017)
[arXiv:1701.03844 [hep-th]].

\bibitem{Viss:2009}
M.~Visser,
arXiv:0912.4757 [hep-th].

\bibitem{Blas:2009}
D.~Blas, O.~Pujolas and S.~Sibiryakov,
Phys. Rev. Lett. \textbf{104}, 181302 (2010)
[arXiv:0909.3525 [hep-th]].

\bibitem{Deve:2018}
D.~O.~Devecioglu and M.~I.~Park,
Phys. Rev. D \textbf{99},
104068 (2019)
[arXiv:1804.05698 [hep-th]].

\bibitem{ONea:2020}
K.~O'Neal-Ault, Q.~G.~Bailey and N.~A.~Nilsson,
Phys. Rev. D \textbf{103},
044010 (2021)
[arXiv:2009.00949 [gr-qc]].

\bibitem{Zhan:2020}
Y.~Zhang \textit{et al.},
Mon. Not. Roy. Astron. Soc. \textbf{501},
1013
(2021)
[arXiv:2007.12607 [astro-ph.CO]].

\bibitem{Park:2009}
M.~I.~Park,
JHEP \textbf{09}, 123 (2009);
JCAP \textbf{01}, 001 (2010)
[arXiv:0906.4275 [hep-th]].

\bibitem{Dutt:2009}
S.~Dutta and E.~N.~Saridakis,
JCAP \textbf{01}, 013 (2010)
[arXiv:0911.1435 [hep-th]].

\bibitem{Son:2010}
E.~J.~Son and W.~Kim,
JCAP \textbf{06}, 025 (2010)
[arXiv:1003.3055 [hep-th]].

\bibitem{Stei:2012}
G.~Steigman,
Adv. High Energy Phys. \textbf{2012}, 268321 (2012)
[arXiv:1208.0032 [hep-ph]].

\bibitem{Nils:2018}
N.~A.~Nilsson and E.~Czuchry,
Phys. Dark Univ. \textbf{23}, 100253 (2019)
[arXiv:1803.03615 [gr-qc]].

\bibitem{Robe:2015} C.~ P.~ Robert, arXiv:1504.01896 [stat.CO].

\bibitem{Dunk:2004}
J.~Dunkley \textit{et al.},
Mon. Not. Roy. Astron. Soc. \textbf{356}, 925
(2005)
[arXiv:astro-ph/0405462 [astro-ph]].

\bibitem{Lewi:2019}
A.~Lewis,
arXiv:1910.13970 [astro-ph.IM].

\bibitem{Zhai:2018}
Z.~Zhai and Y.~Wang,
JCAP \textbf{07}, 005 (2019)
[arXiv:1811.07425 [astro-ph.CO]].

\bibitem{BOSS:2016}
S.~Alam \textit{et al.} [BOSS],
Mon. Not. Roy. Astron. Soc. \textbf{470},
2617
(2017)
[arXiv:1607.03155 [astro-ph.CO]].

\bibitem{Ata:2017}
M.~Ata  \textit{et al.},
Mon. Not. Roy. Astron. Soc. \textbf{473},
4773
(2018)
[arXiv:1705.06373 [astro-ph.CO]].

\bibitem{deSa:2019}
V.~de Sainte Agathe \textit{et al.},
Astron. Astrophys. \textbf{629}, A85 (2019)
[arXiv:1904.03400 [astro-ph.CO]].


\bibitem{Blak:2012}
C.~Blake \textit{et al.},
Mon. Not. Roy. Astron. Soc. \textbf{425}, 405
(2012)
[arXiv:1204.3674 [astro-ph.CO]].

\bibitem{Scol:2017}
D.~M.~Scolnic \textit{et al.},
Astrophys. J. \textbf{859},
101 (2018)
[arXiv:1710.00845 [astro-ph.CO]].

\bibitem{Liu:2014}
J.~Liu and H.~Wei,
Gen. Rel. Grav. \textbf{47},
141 (2015)
[arXiv:1410.3960 [astro-ph.CO]].

\bibitem{Wong:2019}
K.~C.~Wong \textit{et al.},
Mon. Not. Roy. Astron. Soc. \textbf{498},
1420
(2020)
[arXiv:1907.04869 [astro-ph.CO]].

\bibitem{More:2015}
M.~Moresco,
Mon. Not. Roy. Astron. Soc. \textbf{450},
L16
(2015)
[arXiv:1503.01116 [astro-ph.CO]].


\bibitem{Efst:2020}
G.~Efstathiou and S.~Gratton,
Mon. Not. Roy. Astron. Soc. \textbf{496},
L91
(2020)
[arXiv:2002.06892 [astro-ph.CO]].

\bibitem{Gonz:2021}
J.~E.~Gonzalez \textit{et al.},
JCAP \textbf{11},
060 (2021)
[arXiv:2104.13455 [astro-ph.CO]].

\bibitem{Park:2018}
C.~G.~Park and B.~Ratra,
Astrophys. Space Sci. \textbf{364},
134 (2019)
[arXiv:1809.03598 [astro-ph.CO]].

\bibitem{Nune:2020}
R.~C.~Nunes and A.~Bernui,
Eur. Phys. J. C \textbf{80},
1025 (2020)
[arXiv:2008.03259 [astro-ph.CO]].

\bibitem{Beni:2020}
D.~Benisty and D.~Staicova,
Astron. Astrophys. \textbf{647}, A38 (2021)
[arXiv:2009.10701 [astro-ph.CO]].

\bibitem{Vagnozzi:2020}
S.~Vagnozzi \textit{et al.},
Phys. Dark Univ. \textbf{33}, 100851 (2021)
[arXiv:2010.02230 [astro-ph.CO]].

\bibitem{Vagnozzi:2020dfn}
S.~Vagnozzi, A.~Loeb and M.~Moresco,
Astrophys. J. \textbf{908},
84 (2021)
[arXiv:2011.11645 [astro-ph.CO]].

\bibitem{CMB Wiki}
$https://wiki.cosmos.esa.int/planck-legacy-archive/images/2/21/\\
Baseline_params_table_2018_95pc_v2.pdf $

\bibitem{Zhai:2019}
Z.~Zhai \textit{et al.},
JCAP \textbf{07}, 009 (2020)
[arXiv:1912.04921 [astro-ph.CO]].

\bibitem{SPT:2021}
L.~Balkenhol \textit{et al.} [SPT-3G],
Phys. Rev. D \textbf{104},
083509 (2021)
[arXiv:2103.13618 [astro-ph.CO]].

\bibitem{Mang:2001}
G.~Mangano, G.~Miele, S.~Pastor and M.~Peloso,
Phys. Lett. B \textbf{534}, 8
(2002)
[arXiv:astro-ph/0111408 [astro-ph]].

\bibitem{Flor:2020}
M.~M.~Flores and A.~Kusenko,
Phys. Rev. Lett. \textbf{126},
041101 (2021)
[arXiv:2008.12456 [astro-ph.CO]].

\bibitem{Grae:2018}
L.~L.~Graef, M.~Benetti and J.~S.~Alcaniz,
Phys. Rev. D \textbf{99},
043519 (2019)
[arXiv:1809.04501 [astro-ph.CO]].

\bibitem{Jeff:1939}
H.~Jeffreys,
``The Theory of Probability'' (Oxford, Oxford Univ. Press, Englad, 1961).

\bibitem{Kass:1995}
R.~E.~Kass and A.~E.~Raftery,
J. Am. Statist. Assoc. \textbf{90},
773
(1995).

\bibitem{Ness:2012}
S.~Nesseris and J.~Garcia-Bellido,
JCAP \textbf{08}, 036 (2013)
[arXiv:1210.7652 [astro-ph.CO]].

\bibitem{Trot:2008}
R.~Trotta,
Contemp. Phys. \textbf{49}, 71
(2008)
[arXiv:0803.4089 [astro-ph]].

\bibitem{Argu:2015}
C.~Arg\"uelles, N.~Grandi and M.~I.~Park,
JHEP \textbf{10}, 100 (2015)
[arXiv:1508.04380 [hep-th]].

\bibitem{Knox:2019}
L.~Knox and M.~Millea,
Phys. Rev. D \textbf{101},
043533 (2020)
[arXiv:1908.03663 [astro-ph.CO]].

\bibitem{Ries:2020}
A.~G.~Riess \textit{et al.},
Astrophys. J. Lett. \textbf{908},
L6 (2021)
[arXiv:2012.08534 [astro-ph.CO]].


\bibitem{Lewi:1999}
A.~Lewis, A.~Challinor and A.~Lasenby,
Astrophys. J. \textbf{538}, 473
(2000)
[arXiv:astro-ph/9911177 [astro-ph]].

\bibitem{Lesg:2011}
J.~Lesgourgues,
[arXiv:1104.2932 [astro-ph.IM]].

\bibitem{Wang:2015}
Y.~Wang and M.~Dai,
Phys. Rev. D \textbf{94},
083521 (2016)
[arXiv:1509.02198 [astro-ph.CO]].

\bibitem{Hu:1995}
W.~Hu and N.~Sugiyama,
Astrophys. J. \textbf{471}, 542-570 (1996)
[arXiv:astro-ph/9510117 [astro-ph]].

\bibitem{Eise:1997}
D.~J.~Eisenstein and W.~Hu,
Astrophys. J. \textbf{496}, 605 (1998)
[arXiv:astro-ph/9709112 [astro-ph]].

\bibitem{SNLS:2011}
A.~Conley \textit{et al.} [SNLS],
Astrophys. J. Suppl. \textbf{192}, 1 (2011)
[arXiv:1104.1443 [astro-ph.CO]].

\bibitem{Hui:2005}
L.~Hui and P.~B.~Greene,
Phys. Rev. D \textbf{73}, 123526 (2006)
[arXiv:astro-ph/0512159 [astro-ph]].

\bibitem{Wang:2013}
Y.~Wang and S.~Wang,
Phys. Rev. D \textbf{88},
043522 (2013)
[Erratum: Phys. Rev. D \textbf{88},
069903 (2013)]
[arXiv:1304.4514 [astro-ph.CO]].

\bibitem{Suyu:2016}
S.~H.~Suyu \textit{et al.}
Mon. Not. Roy. Astron. Soc. \textbf{468},
2590
(2017)
[arXiv:1607.00017 [astro-ph.CO]].

\bibitem{Jime:2001}
R.~Jimenez and A.~Loeb,
Astrophys. J. \textbf{573}, 37
(2002)
[arXiv:astro-ph/0106145 [astro-ph]].

\bibitem{More:2016}
M.~Moresco \textit{et al.},
JCAP \textbf{05}, 014 (2016)
[arXiv:1601.01701 [astro-ph.CO]].




\end{thebibliography}
\end{document}